\documentclass[aps,prd,twocolumn,nofootinbib,showpacs,superscriptaddress,floatfix]{revtex4-1}  
\usepackage{graphicx}  
\usepackage{dcolumn}   
\usepackage{bm}        
\usepackage{amssymb}   
\usepackage{amsmath}
\usepackage{comment}
\usepackage[hidelinks]{hyperref}
\usepackage{graphicx}
\usepackage{hyperref}
\usepackage{subfig}
\usepackage{color}
\usepackage{slashed}
\usepackage{url}
\usepackage{hyperref}
\usepackage{lipsum}

\newcommand{\kt}{{\bf k}}
\newcommand{\Bt}{{\bf B}}
\newcommand{\vt}{{\bf v}}
\newcommand{\xt}{{\bf x}}
\newcommand{\et}{{\bf e}}

\newcommand{\gsph}{{\gamma_{\rm sph}}}
\newcommand{\Gsph}{{\Gamma_{\rm sph}}}
\newcommand{\geta}{{\gamma_{\eta}}}
\newcommand{\kkc}{{\left(\frac{ |{\bf k}| \cos\theta_{\kt\Bt}}{k_{\rm CMW}}\right)}}
\newcommand{\kc}{{k_{\rm CMW}}}
\newcommand{\eb}{{eB/T^2}}

\begin{document}
 
\author{Lillian de Bruin}
\email{debruin@thphys.uni-heidelberg.de}
\affiliation{Physics and Astronomy Department, Stony Brook University, Stony Brook, NY 11794, USA}
\affiliation{Institute for Theoretical Physics, Heidelberg University, Philosophenweg 12, 69120 Heidelberg, Germany}
\author{S\"{o}ren Schlichting}
\affiliation{Fakult\"{a}t f\"{u}r  Physik, Universit\"{a}t Bielefeld, D-33615 Bielefeld, Germany}

\title{Sphaleron damping and effects on vector and axial charge transport in high-temperature QCD plasmas}

\begin{abstract}
We modify the anomalous hydrodynamic equations of motion to account for dissipative effects due to QCD sphaleron transitions. By investigating the linearized hydrodynamic equations, we show that sphaleron transitions lead to nontrivial effects on vector and axial charge transport phenomena in the presence of a magnetic field. Due to the dissipative effects of sphaleron transitions, a wavenumber threshold $k_{\rm CMW}$ emerges characterizing the onset of chiral magnetic waves. Sphaleron damping also significantly impacts the time evolution of both axial and vector charge perturbations in a QCD plasma in the presence of a magnetic field. Based on our analysis of the linearized hydrodynamic equations, we also investigate the dependence of the vector charge separation on the sphaleron transition rate, which may have implications for the experimental search for the Chiral Magnetic Effect in Heavy Ion Collisions.

\end{abstract}
\date{\today}

\maketitle


\section{Introduction}\label{sec:intro}

Chiral transport phenomena have recently attracted a significant amount of attention in experimental and theoretical studies, as they may have a significant impact on the collective dynamics of systems possessing (approximately) chiral fermions. Since such systems are ubiquitous in nature, possible manifestations of these phenomena occur across a diverse range of energy scales, with examples including the dynamics of baryo- and magnetogenesis in the early universe \cite{Brandenburg:2017rcb}, the quark-gluon plasma (QGP) in heavy ion collisions \cite{Skokov:2016yrj}, and Dirac and Weyl semimetals in condensed matter systems \cite{Li:2014bha}. 

Unlike ordinary transport phenomena, which describe the macroscopic dynamics of conserved energy-momentum and (vector) charges on large time and distance scales, novel chiral transport phenomena are linked to the dynamics of axial charges, which generically are not conserved due to quantum anomalies \cite{Bell:1969ts,Adler:1969gk}. Despite the expected importance of axial charge changing processes in high temperature QCD plasmas, the effects of such processes for describing anomalous transport phenomena in heavy-ion collisions are frequently neglected in phenomenological studies~\cite{Horvath:2019dvl}, and have only been explored to a limited extent in the studies of~\cite{Stephanov:2014dma,Jimenez-Alba:2014iia}. The primary objective of this paper is to clarify under which conditions anomalous charge transport in high-temperature QCD plasmas can be described macroscopically by anomalous relativistic hydrodynamics, and to explore the extent to which the non-conservation of axial charge due to QCD sphaleron transitions affects transport processes in a QCD plasma. 

Starting with a general discussion of axial charge dynamics in high-temperature QCD plasmas in Sec.~\ref{sec:chiraldyn}, we establish the conditions under which a macroscopic description can be justified, and subsequently in Sec.~\ref{sec:hydrotings} demonstrate how to include axial charge changing processes due to sphaleron transitions in the anomalous hydrodynamic description of high-temperature QCD plasmas. Based on this framework, we demonstrate in Sec.~\ref{sec:excitations} that sphaleron transitions have a non-trivial effect on the coupled hydrodynamic behavior of axial and vector charges in the presence of a magnetic field. Strikingly, we observe that the inclusion of the sphaleron damping term leads to the emergence of a wavenumber threshold that characterizes the hydrodynamic behavior of coupled charge modes and indicates the formation of Chiral Magnetic Waves (CMWs). Due to the particular form of the chiral anomaly, the dissipative effects due to sphaleron transitions also induce a non-trivial coupling between different species of chiral fermions, which we discuss using the example of the $u,d$ light flavor sector of QCD. Subsequently, in Sec.~\ref{sec:diffusion}, we investigate the sensitivity of axial and vector charge transport in the presence of a magnetic field to the sphaleron transition rate. We provide illustrative examples of vector and axial charge separation by numerically solving the linearized hydrodynamic equations, as well as an analytic expression for the vector charge separation in a space-time homogeneous plasma that elucidates its dependence on the sphaleron rate. We finally conclude in Sec.~\ref{sec:conclusion} with a summary of our findings and comments on the implications for the experimental search for chiral transport phenomena in heavy-ion collisions. 



\section{Chirality charge dynamics in high-temperature QCD plasmas}\label{sec:chiraldyn}

Specifically, for an $SU(N_c) \times U(1)$ gauge theory coupled to $N_f$ flavors of massless Dirac fermions, which describes a high temperature QCD plasma in the presence of electromagnetic fields, the non-conservation of the axial current $j^{\mu}_{A,f}(x)=\bar{\psi}_{f}(x) \gamma^{5}\gamma^{\mu} \psi_{f}(x)$ of each fermion flavor takes the form of a local balance equation
\begin{align}
\partial_{\mu}j_{A,f}^{\mu}(x)&=-\frac{(eq_f)^2 N_{c}}{8\pi^2}F_{\mu\nu}(x)\tilde{F}^{\mu\nu}(x)\nonumber\\
&-\frac{g^2}{16\pi^2}G_{\mu\nu}^{a}(x)\Tilde{G}_{a}^{\mu\nu}(x), 
\label{eq:abj-anomaly}
\end{align}
where $e,g$ are the $U(1)$ and $SU(N_c)$ gauge couplings, $F_{\mu\nu}$ and $G_{\mu\nu}^{a}$ denote the corresponding Abelian and non-Abelian field strength tensors, $\Tilde{F}^{\mu\nu}=\frac{1}{2}\epsilon^{\mu\nu\alpha\beta}F_{\alpha\beta}$ and $\Tilde{G}^{\mu\nu}_{a}=\frac{1}{2}\epsilon^{\mu\nu\alpha\beta}G_{\alpha\beta}^{a}$, are their duals and $q_f$ is the electric charge of each fermion flavor. By recognizing the terms on the r.h.s. of Eq.~(\ref{eq:abj-anomaly}) as the covariant divergence of the respective Chern-Simons currents,
$\partial_{\mu}Q^{\mu}(x)=\frac{e^2}{16\pi^2} F_{\mu\nu}(x)\tilde{F}^{\mu\nu}(x)$, $\partial_{\mu}K^{\mu}(x)=\frac{g^2}{32\pi^2} G_{\mu\nu}^{a}(x)\tilde{G}^{\mu\nu}_{a}(x)$
 the axial anomaly relation in Eq.~(\ref{eq:abj-anomaly}) expresses the local conservation of the overall chirality of fermions $j^{0}_{A,f}$ and gauge fields $2q_{f}^{2}N_{c}Q^{0},2K^{0}$ for each massless flavor. Since anomalous transport phenomena such as the chiral magnetic effect (CME)~\cite{Fukushima:2008xe} only occur when a net chirality imbalance is present in the fermion sector $(j^{0}_{A,f} \neq 0)$, it is thus important to understand how a chiral charge imbalance is transferred and re-distributed between fermions and gauge fields on the macroscopic time scales of interest. 

Due to their expected importance, different mechanism of chirality transfer have been explored in the context of condensed matter physics~\cite{Gorbar:2014qta}, nuclear physics~\cite{Schlichting:2022fjc}, and cosmology~\cite{Figueroa:2019jsi}. Straightforwardly, in QED plasmas or QED-like materials, a chiral charge imbalance in the fermion sector can be created via the application of (aligned) external electric and magnetic fields \cite{Li:2014bha}, while spacetime-dependent fluctuations of (chromo-)electromagnetic fields \cite{Mace:2016svc,Figueroa:2019jsi} can generate local fluctuations of the chiral charge imbalance of fermions in QED and QCD plasmas. Conversely, a chirality imbalance in the fermion sector can generate chiral plasma instabilities in both QED and QCD plasmas \cite{Akamatsu:2013pjd,Hirono:2015rla}, which induce a transfer of chirality from fermions to gauge fields. However, on sufficiently large time and distance scales, the transfer of chirality in non-Abelian gauge theories, such as QCD, is believed to be dominated by so-called sphaleron transitions between different topological sectors of the $SU(N_c)$ gauge fields~ \cite{McLerran:1990de,Moore:2010jd,Mace:2016svc,Schlichting:2022fjc}.

By virtue of the non-trivial topology of the $SU(N_{c})$ gauge field configurations in the physical real space $\mathbb{R}^{3} \cup \{\infty\}$, non-Abelian gauge theories such as QCD feature an infinite number of topologically inequivalent but otherwise degenerate field configurations labelled by an integer Chern-Simons number $N_{CS}=\int d^3\mathbf{x}~K^{0}(\mathbf{x})$. In high-temperature QCD plasmas, transitions between different topological sectors are thermally activated by finite-energy configurations called sphalerons. Sphaleron transitions between different topological sectors result in a change in $N_{CS}$ by plus/minus unity, which according to Eq.~(\ref{eq:abj-anomaly}) results in a change of the net-axial charge of fermions $J_{A,f}^{0}=\int d^3\mathbf{x}~J_{A}^{0}(\mathbf{x})$ by plus/minus two units for each flavor. While in charge-neutral plasmas the dynamics of sphaleron transitions thus induces time dependent fluctuations of the chiral charge imbalance $J_{0}^{A}=\sum_{f} j_{A,f}^{0}$ of fermions, sphaleron transitions in a chirally imbalanced plasma $(J_{A}^{0}\neq 0)$
exhibit a bias towards erasing any pre-existing charge imbalance $J_{0}^{A}$~\cite{McLerran:1990de,Schlichting:2022fjc}, such that on asymptotically large time scales any chiral charge imbalance of fermions will disappear. Since a non-vanishing chiral charge imbalance is however required to realize e.g. the chiral magnetic effect, one concludes that anomalous transport phenomena in high-temperature QCD plasmas are in a sense intrinsically non-equilibrium phenomena, which can only occur on transient time scales before the chiral charge imbalance is eventually erased.

Evidently, the typical time scale for which a chiral charge imbalance can persist in a high-temperature QCD plasma then crucially depends on the rate of sphaleron transitions. Despite the fact that sphaleron transitions are known to occur in high-temperature QCD plasmas, it is notoriously challenging to compute the sphaleron rate from first principles \cite{Arnold:1987zg,Moore:2010jd}. The sphaleron transition rate is defined as the thermal expectation value of the zero frequency, zero momentum limit of the Wightman correlation function of $G_{\partial_{\mu}K^{\mu}\partial_{\mu}K^{\mu}}(\omega,\mathbf{k})$ as
\begin{align}
\Gsph&= \int d^4X\bigg\langle\frac{g^2}{32\pi^2} G_{\mu\nu}^a\Tilde{G}_{a}^{\mu\nu}(X)\frac{g^2}{32\pi^2}G_{\alpha\beta}^a\Tilde{G}_{a}^{\alpha\beta}(0)\bigg\rangle\ \
\end{align}
and describes the occurrence of a Chern-Simons number-changing process per unit volume per unit time~\cite{Moore:2010jd}. Evaluated at weak coupling for $SU(N_c)$ gauge theories, the sphaleron transition rate is parametrically given by $\Gsph\propto \alpha_S^5T^4$, where $\alpha_S=g^2/4\pi$~\cite{Moore:2010jd}, while at strong coupling, the rate of sphaleron transitions is computed via the AdS/CFT correspondence for an $\mathcal{N}=4$ Supersymmetric Yang-Mills plasma is given by $N\rightarrow\infty$: $\Gsph=(g^2N)^2T^4/256\pi^3$\cite{Basar:2012gh}. Very recently, (quenched) lattice QCD calculations~\cite{Altenkort:2020axj} have determined the sphaleron transition rates at temperatures $T=1.5T_{c}$ to be on the order of $\Gamma_{\rm sph}=(0.02 - 0.2) T^{4}$, with large systematic uncertainties stemming from the analytic continuation of Euclidean correlation functions to Minkowski space. Even though the estimated rates from lattice QCD are actually sizeable, we will demonstrate shortly that a more careful assessment of their magnitude suggests that an effective macroscopic description of axial charge transport in high-energy heavy-ion collisions may still be warranted.

\section{Hydrodynamic description of anomalous transport in QCD-like theories}\label{sec:hydrotings}
Although chiral transport phenomena in high-temperature QCD plasmas are in principle intrinsically non-equilibrium phenomena, their possible macroscopic manifestations also emerge naturally within the framework of anomalous hydrodynamics \cite{Son:2009tf}. Indeed, if the process of axial charge equilibration is slow compared to the typical kinetic equilibration of the QCD plasma, the axial currents $j_{A,f}^{\mu}$ represent additional slow variables whose dynamics can be described macroscopically by introducing additional axial chemical potentials $\mu_{A}^{f}$ associated with the residual deviations of the axial charge $j_{A,f}^{0}$ from the genuine equilibrium state. However, a meaningful hydrodynamic description based on an expansion around transient equilibrium states with non-vanishing axial chemical potentials ($\mu_{A}^{f}\neq0$) can only be achieved if the equilibration of axial charge is slow compared to the equilibration of the system. Certainly this is the case for weakly-coupled $SU(N_c)$ plasmas, where the timescale of axial charge relaxation due to sphaleron transitions $\tau_{\rm sph}\approx \frac{\chi_{A} T}{\Gamma_{\rm sph}} \sim \alpha_S^{-5}T^{3}$~\cite{Moore:2010jd} is much larger than the timescale associated with the kinetic equilibration of the plasma, $\tau_{\rm kin}\approx \frac{4\pi \eta/s}{T} \sim\alpha_S^{-2}T^{-1}$ \cite{Arnold:2000dr}. When considering the QGP created in heavy ion collisions at RHIC and LHC energies, where temperatures typically range up to $\sim 4 T_{c}$, one finds that with the estimate of $\Gamma_{\rm sph} \approx 0.1 T^4$ from \cite{Altenkort:2020axj} $\tau_{\rm sph} \sim 10 T^{-1}$ can be larger, but not significantly larger than $\tau_{\rm kin} \approx 2 T^{-1} {\rm fm/c}$ for favorable values of the transport coefficient $\eta/s=0.16$~\cite{Bernhard:2019bmu}.

Now that we have established the anticipated range of applicability of an effective macroscopic description, we proceed to develop the hydrodynamic description of anomalous charge transport following previous works~\cite{Son:2009tf,Sadofyev:2010pr}. We consider a viscous relativistic fluid in $3+1$ spacetime dimensions, governed by the Minkowski metric $g^{\mu\nu}=(-1,{\bf 1})$, with conserved $U(1)$ vector currents\footnote{Note that the vector current is defined such that the electric current is $j_{\rm el,f}^{\mu}=eq_fj_{V,f}^{\mu}$.} $j_{V,f}^{\mu}=\langle\bar{\Psi}_{f}\gamma^{\mu}\Psi_{f} \rangle$  and $U(1)$ axial currents $j_{A,f}^{\mu}=\langle\bar{\Psi}_{f}\gamma^{\mu}\gamma_5\Psi_{f}\rangle$ that are not conserved due to the axial anomaly for $N_f$ flavors of massless Dirac fermions. In the presence of a slowly-varying, non-dynamical background electromagnetic field, the conservation laws take the form
\begin{align}
\partial_{\mu}T^{\mu\nu}&=\sum_{f} eq_fF^{\nu\lambda}j_{\lambda,f}^V, \label{eq:cons-en}\\
\partial_{\mu}j_{V,f}^{\mu}&= 0,\label{eq:cons-vec}\\
\partial_{\mu}j_{A,f}^{\mu}&= (eq_f)^2CE^{\mu}B_{\mu}-\frac{g^2}{16\pi^2} G_{\mu\nu}^{a}\tilde{G}^{\mu\nu}_{a}, \label{eq:cons-ax}
\end{align}
where the right hand side of Eq. (\ref{eq:cons-en}) reflects work done on the system by the external electromagnetic field. Conversely, the right hand side of Eq. (\ref{eq:cons-ax}) reflects the non-conservation of axial charge, where effects due to the Abelian chiral anomaly are described explicitly by the term $(eq_f)^2CE^{\mu}B_{\mu}$ with the anomaly coefficient $C=N_c/2\pi^2$. Non-Abelian contributions to the axial anomaly are described by the last term in Eq.~(\ref{eq:cons-ax}), which -- in accordance with the discussion in Sec.~\ref{sec:chiraldyn} -- tend to erase any pre-existing axial charge imbalance. By following the arguments of Shaposnikov, McLerran, and Mottola~\cite{McLerran:1990de}, the expectation value of $\left \langle \frac{g^2}{16\pi^2} G_{\mu\nu}^{a}\tilde{G}^{\mu\nu}_{a} \right \rangle$ can be expressed in terms of the sphaleron transition rate $\Gsph$ as
\begin{align}
\left \langle \frac{g^2}{16\pi^2} G_{\mu\nu}^{a}\tilde{G}^{\mu\nu}_{a} \right \rangle = 4 \Gamma_{\rm sph} \sum_{f} \frac{\mu_{f,A}}{T}\;,
\end{align}
which in the presence of finite axial chemical potentials $\sum_{f} \mu_{f,A}$ is manifestly non-zero.
We note that even though individual sphaleron transitions represent singular local events, which result in an integer change of the net axial charge, the macroscopic description in Eq.~(\ref{eq:cons-ax}) is valid over large time and distance scales, where on average multiple sphaleron transitions provide a dissipative effect on the axial charge evolution of the fluid.

The electromagnetic fields are defined in Lorentz covariant form,
\begin{align}
E^{\mu}=F^{\mu\nu}u_{\nu},\qquad B^{\mu}=\frac{1}{2}\epsilon^{\mu\nu\alpha\beta}u_{\nu}F_{\alpha\beta},
\end{align}
such that $u_{\mu}B^{\mu}=u_{\mu}E^{\mu}=0$. Here, $u^{\mu}$ denotes the rest-frame velocity field, which, following Landau and Lifshitz, is defined as the timelike eigenvector of the energy momentum tensor: $-u_{\mu}T^{\mu\nu}=\epsilon u^{\nu}$ such that $u^2=-1$. Besides $u^{\mu}$, we use local temperature $T$ and chemical potentials $\mu_{V_f/A)f}$ for each fermion flavor as thermodynamic variables. We also define the vorticity, 
\begin{align}
\omega^{\mu}=\frac{1}{2}\epsilon^{\mu\nu\alpha\beta}u_{\nu}\partial_{\alpha}u_{\beta},
\end{align}
which must be included in the hydrodynamic description of anomalous relativistic fluids \cite{Son:2009tf}. 

Next, to obtain the complete set of hydrodynamic equations of the system, we supplement Eqs. (\ref{eq:cons-en})-(\ref{eq:cons-ax}) with the constitutive relations for the vector/axial currents $j^{\mu}_{V/A,f}$ and the energy momentum tensor $T^{\mu\nu}$, which, in the most general form in the Landau frame, are written as
\begin{align}
T^{\mu\nu}&=(\epsilon+P)u^{\mu}u^{\nu}+Pg^{\mu\nu}+\tau^{\mu\nu}, \label{eq:tmunu}\\
j_{V,f}^{\mu}&=n_{V,f}u^{\mu}+\nu_{V,f}^{\mu}, \label{eq:jvec} \\
j_{A,f}^{\mu}&=n_{A,f}u^{\mu}+\nu_{A,f}^{\mu}, \label{eq:jax} 
\end{align} 
where $\epsilon=\epsilon(T,\mu_{V,f},\mu_{A,f})$ denotes the energy density and $P=P(T,\mu_{V,f},\mu_{A,f})$ is the thermodynamic pressure.
By definition, $\tau^{\mu\nu}$ satisfies the relation $u_{\mu}\tau^{\mu\nu}=0$, and the dissipative currents $\nu^{\mu}_{V/A,f}$ are defined such that $u_{\mu}\nu^{\mu}_{V/A,f}=0$ and $n_{V/A,f}=u_{\mu}j_{V/A,f}^{\mu}$ is the vector/axial charge density in the local rest frame of the fluid. We also note that, following the common practice in the field of heavy-ion physics, we take all of the above quantities to correspond to their expectation values, and we will not consider thermodynamic fluctuations in this study.

In this study, we restrict ourselves to studying the dissipative corrections $\tau^{\mu\nu}$ and $\nu_{V/A}^{\mu}$ up to first order in gradients of the hydrodynamic variables and external fields. The correction to Eq. (\ref{eq:tmunu}) is then given by
 \begin{align}
 \tau^{\mu\nu}=-\eta\sigma^{\mu\nu}-\zeta\Delta^{\mu\nu}\partial\cdot u,
 \label{eq:taumunu}
 \end{align}
which corresponds to ordinary first-order viscous corrections to energy-momentum transport, where $\eta$ and $\zeta$ are the shear and bulk viscosity,  $\sigma^{\mu\nu}\equiv\Delta^{\mu\alpha}\Delta^{\nu\beta}\left(\partial_{\alpha}u_{\beta}+\partial_{\beta}u_{\alpha}-\frac{2}{3}g_{\alpha\beta}\partial\cdot u\right)$ is the transverse traceless symmetric shear-stress tensor, and $\Delta^{\mu\nu}=g^{\mu\nu}+u^{\mu}u^{\nu}$ denotes the spatial projector orthogonal to the direction of fluid flow. Similarly, the first order viscous corrections to Eqs. \eqref{eq:jvec} and \eqref{eq:jax} take the following general form
\begin{align}
\nu_{V,f}^{\mu}=&-\sigma_{VV}^{ff'}\left(T\Delta^{\mu\nu}\partial_{\nu}\frac{\mu_{V,f'}}{T}-eq_{f'}E^{\mu}\right) \label{eq:nuv} \\ 
& -\sigma_{VA}^{ff'}T\Delta^{\mu\nu}\partial_{\nu}\frac{\mu_{A,f'}}{T}
 +eq_{f}\sigma_{VB}^{f}B^{\mu}+\xi_{V,f}\omega^{\mu},  \nonumber\\
\nu_{A,f}^{\mu}=&-\sigma_{AV}^{ff'}\left(T\Delta^{\mu\nu}\partial_{\nu}\frac{\mu_{V,f'}}{T}-eq_{f'}E^{\mu}\right) \label{eq:nua} \\ 
& -\sigma_{AA}^{ff'}T\Delta^{\mu\nu}\partial_{\nu}\frac{\mu_{A,f'}}{T} 
 +eq_{f}\sigma_{AB}^{f}B^{\mu}+\xi_{A,f}\omega^{\mu}, \nonumber
\end{align}
and, if not stated otherwise, we will consider the various conductivity tensors to be diagonal in flavor space $\sigma^{ff'}=\sigma \delta^{ff'}$, to comply with the $SU(N_{f}) \times SU(N_{f})$ flavor symmetry of a charge-neutral plasma in the chirally symmetric phase. We note that the various coefficients in Eqs. \eqref{eq:nuv} and \eqref{eq:nua} have straightforward physical interpretations, where $\sigma_{VV}$ and $\sigma_{AA}$ are the vector and axial conductivities, while the ``off-diagonal'' transport coefficients $\sigma_{VA}$ and $\sigma_{AV}$ describe the coupled transport of axial and vector charges associated with the chiral electric separation effect \cite{Huang:2013iia}, with $\sigma_{VA}=\sigma_{AV}$ due to the Onsager relations \cite{Onsager:1931jfa}. The other coefficients are related to anomalous chiral transport phenomena associated with the magnetic field and vorticity: $\sigma_{BV}$ is the conductivity due to the chiral magnetic effect \cite{Fukushima:2008xe}, $\sigma_{BA}$ is the conductivity due to the chiral separation effect \cite{Son:2004tq,Metlitski:2005pr}, $\xi_V$ is the coupling of the chiral vortical effect, and $\xi_A$ is the spin-vorticity coupling\cite{Son:2009tf}.

Evidently, the transport coefficients in Eqs. \eqref{eq:taumunu}-\eqref{eq:nua} are constrained by the second law of thermodynamics, which requires local entropy production to be non-negative. Based on this requirement, it follows directly that the ordinary transport coefficients satisfy the relations $\eta\geq0$ and $\zeta\geq0$ for the shear and bulk viscosities in Eq. \eqref{eq:taumunu}, as well as $\Gamma_{\rm sph} \geq 0$, $\sigma_{VV}\geq0$ and $\sigma_{AA}\geq0$, while $\sigma_{AV}\sigma_{VA}\leq\sigma_{AA}\sigma_{VV}$, for the sphaleron rate and the various conductivities in Eqs. \eqref{eq:nuv} and \eqref{eq:nua}. Strikingly, as pointed out in a seminal paper by Son and Surowka~\cite{Son:2009tf} and follow-up works~\cite{Sadofyev:2010pr}, the various anomalous chiral transport coefficients in Eqs. \eqref{eq:taumunu}-\eqref{eq:nua} are constrained to an even greater extent by the same condition. To show this, we quantify entropy production via the entropy current,
\begin{align}
S^{\mu}&=& su^{\mu}+D_BB^{\mu}+D_{\omega}\omega^{\mu}-\frac{\mu_{V,f}}{T}\nu_{V,f}^{\mu}-\frac{\mu_{A,f}}{T}\nu_{A,f}^{\mu}, \nonumber \label{eq:entcurrent} \\
\end{align}
where $s$ is defined by the thermodynamic relation $Ts=(\epsilon+P)-\mu_{V,f}n_{V,f}-\mu_{A,f}n_{A,f}$ and $D_{B,\omega}$ are general functions of temperature $T$ and chemical potentials $\mu_{V/A,f}$. By exploiting the hydrodynamic equations and thermodynamic relations, we can then express the divergence of $S^{\mu}$ as (see  Appendix \ref{app:divergences} for details)
\begin{align}
&
\partial_{\mu}S^{\mu}=-\frac{1}{T}\partial_{\mu}u_{\nu}\tau^{\mu\nu} -\nu_{V,f}^{\mu}\left(\partial_{\mu}\frac{\mu_{V,f}}{T}-\frac{eq_f}{T}E_{\mu}\right) \nonumber \\
& -\nu_{A,f}^{\mu}\partial_{\mu}\frac{\mu_{A,f}}{T}+ 4\Gsph\left(\sum_{f} \frac{\mu_{A,f}}{T} \right)^2\nonumber\\
& -\left( \sum_{f} \frac{\mu_{A,f}}{T} (eq_f)^2 \right)CE^{\mu}B_{\mu}+\partial_{\mu}\left(D_BB^{\mu}+D_{\omega}\omega^{\mu}\right), \nonumber\\\label{eq:divs}
\end{align}
and require positive entropy production with the condition 
\begin{align}
\partial_{\mu}S^{\mu}\geq0.
\end{align}
Dissipative effects due to shear $(\eta)$ and bulk $(\zeta)$ viscous corrections, vector and axial charge diffusion $(\sigma_{V/A V/A})$, as well as sphaleron damping $(\Gamma_{\rm sph})$ contribute positively to entropy production. Deferring the details of the calculation to Appendix \ref{app:transport}~(see also \cite{Son:2009tf}), one finds that a thermodynamically consistent description of the anomalous transport phenomena associated with the coefficients $\sigma_{VB},\sigma_{BA},\xi_{A},\xi_{V}$, requires these phenomena to be non-disspiative in the sense that their contribution to $\partial_{\mu}S^{\mu}$ vanishes identically. Based on this requirement, following the calculations in \cite{Son:2009tf,Sadofyev:2010pr}, one obtains the following constraints on the anomalous transport coefficients in the single-flavor case:
\begin{align}
\label{eq:Conductivities}
\sigma_{VB} =&\ C \left(\mu_A -\frac{n_V \mu_A \mu_V}{\epsilon+P} \right),\\ 
\sigma_{AB} =&\ C \left(\mu_V - \frac{n_A \mu_A \mu_V}{\epsilon + P}\right) + (e q_f)^{-1} \frac{\partial}{\partial\overline{\mu}_A}g(\overline{\mu}_A),\\
\xi_{A} =&\ C \left(\mu_V^2 - \frac{n_A \mu_A \mu_V^2}{\epsilon +P}\right) +(e q_f)^{-1} \frac{\mu_V}{T}\frac{\partial}{\partial \overline{\mu}_A}g(\overline{\mu}_A) \nonumber\\
&+ \frac{\partial}{\partial \overline{\mu}_A}G(\overline{\mu}_A), \\
\xi_{V} =&\ 2C \left(\mu_V\mu_A - \frac{n_V \mu_A \mu_V^2}{\epsilon +P}\right) + (e q_f)^{-1} g(\overline{\mu}_A),
\end{align}
where $\overline{\mu}_A\equiv\mu_A/T$ and $g$ and $G$ are hitherto arbitrary functions of $\overline{\mu}_A$. These coefficients agree with the single flavor calculations by \cite{Son:2009tf,Sadofyev:2010pr} and with the conductivities calculated microscopically in the original works of \cite{Fukushima:2008xe,Son:2009tf,Kharzeev:2010gd,Son:2004tq,Metlitski:2005pr,Sadofyev:2010pr}. While the positivity of entropy production alone does not lead to such stringent constraints in the multiflavor case (see App.~\ref{app:transport}), we will assume that individual quark flavors behave independently with respect to the chiral anomaly and entropy production and employ the same transport coefficients for the multiflavor case for respective quark flavors.  


Next, we take these coefficients and insert them into the first order corrections to the constitutive relations in Eqs. \eqref{eq:nuv} and \eqref{eq:nua}. We can then take the constitutive relations with the conservation equations to obtain the closed set of hydrodynamic equations that govern the vector and axial charge dynamics in a high-temperature QCD plasma.

\section{Hydrodynamic excitations in charge-neutral plasma}\label{sec:excitations}

Now that we have established the effective macroscopic description of vector and axial charge transport in the presence of QCD sphaleron transitions, we will study the behavior of hydrodynamic excitations on a static equilibrium background, characterized by a fluid velocity field $u^{\mu}=(1,{\bf 0})$, temperature $T$, and vanishing vector/axial charge chemical potentials $\mu_{V_f}=\mu_{A_f}=0$, which is typical in high energy heavy ion collisions. In order to analyze the hydrodynamic equations, we first perform a spatial Fourier transform of the equations of motion, according to
\begin{align}
u^i(t,\xt)&=\int\frac{d^3k}{(2\pi)^3}e^{i\kt\cdot\xt}u_{\kt}^i(t,\kt),
\label{eq:spatial-ft}
\end{align}
and similarly for the other fields, then subsequently linearize the equations of motion around the static equilibrium background. In the presence of an external magnetic field $\Bt$, the velocity field can be decomposed as
\begin{align}
u^{i}=u_{\kt} \frac{\kt^{i}}{|\kt|} + u_{\Bt} \frac{\Bt^{i}}{|\Bt|} + u_{\kt \times \Bt} \frac{\kt \times \Bt}{|\kt \times \Bt|},
\end{align}
such that the longitudinal and transverse components of the fluid velocity fields are given by
\begin{align}
u_{L}&=\frac{\kt^{i}}{|\kt|} u^{i}  = u_{\kt} + u_{\Bt} \cos(\theta_{\kt\Bt}),\\  
u_{\bot}^{i} &= u_{\Bt} \left( \frac{\Bt^{i}}{|\Bt|} - \cos(\theta_{\kt\Bt}) \frac{\kt^{i}}{|\kt|} \right) + u_{\kt \times \Bt} \frac{(\kt \times \Bt)^i}{|\kt \times \Bt|},\quad 
\end{align}
and the transverse component can be further decomposed into two components:
\begin{align}
u_{\perp\Bt}&=  u_{\Bt} \left( \frac{\Bt^{i}}{|\Bt|} - \cos(\theta_{\kt\Bt}) \frac{\kt^{i}}{|\kt|} \right),\\
u_{\perp\perp}&=  u_{\kt \times \Bt} \frac{(\kt \times \Bt)^i}{|\kt \times \Bt|}.
\end{align}

We choose the hydrodynamic variables to be fluctuations in energy density $\delta\epsilon = \delta T^{00}$, momentum density $\pi^i = \delta u^i(\epsilon +P) = \delta T^{0i}$, and charge densities $\delta n_{V,f}= \delta j_{V,f}^0$ and $\delta n_{A,f} = \delta j_{A,f}^0$, as these quantities can be defined microscopically in the underlying theory of QCD. By using thermodynamic relations, it is straightforward to express intensive variables from $T$, $u^{\mu}$, and $\mu_{V/A}$ in terms of extensive ones; in particular we can express changes in charge density in terms of changes in chemical potential according to
\begin{align}
\delta n_{i,f}=\chi_{ij}^{ff'}\delta\mu_{j,f'}\;,\qquad \chi_{ij}^{ff'}=\left(\frac{\partial^2 P}{\partial\mu_{i,f}\partial\mu_{j,f'}}\right)_T,
\end{align}
while changes of the pressure are determined by the equation of state as $\delta P= c_S^2\delta \epsilon$. Since we are considering a charge-neutral background, all transport coefficients are evaluated at $\mu_{V,f}=\mu_{A,f}=0$. We also assume $SU(N_f)\times SU(N_f)$ flavor symmetry, such that $\chi_{ij}^{ff'}=\chi_{i} \delta_{ij} \delta^{ff'}$. We note that in this situation, the ``off-diagonal'' transport coefficients, $\sigma_{VA}=\sigma_{AV}$ in Eqs. \eqref{eq:nuv} and \eqref{eq:nua} also vanish, since the leading-order contributions are $\propto\mu_{A,f}\mu_{V,f}$ \cite{Huang:2013iia}, i.e. of second order in the chemical potentials.

By imposing these conditions, we obtain the complete system of linearized hydrodynamic equations: 
\begin{widetext}
\begin{align}
\partial_t \delta\epsilon + i|\textbf{k}|\pi_L &= 0, \label{eq:sound1}\\
\partial_t \pi_L + i|\textbf{k}| c_s^2 \delta \epsilon + \frac{4}{3}\geta \textbf{k}^2 \pi_L &= 0, \label{eq:sound2} \\
\partial_t \pi_{\perp B} + \geta \textbf{k}^2\pi_{\perp B} &= 0, \label{eq:shear1} \\
\partial_t \pi_{\perp\perp} + \geta \kt^2\pi_{\perp\perp}-\sum_{f} i eq_f |\kt\times\Bt|\left(D_V^{f}\delta n_{V,f}\right) &=0, \label{eq:shear2} \\
\partial_{t}\delta n_{V,f} + D_V\textbf{k}^2 \delta n_{V,f} + e q_f C\frac{i\textbf{k}\cdot\textbf{B}}{\chi_A}\delta n_{A,f} &= 0, \label{eq:nV} \\
\partial_t \delta n_{A,f} + D_A\textbf{k}^2 \delta n_{A,f}  + e q_f C\frac{i\textbf{k}\cdot\textbf{B}}{\chi_V}\delta n_{V,f}&= - \gsph \sum_{f} \delta n_{A,f},\qquad\label{eq:nA}
\end{align}
\end{widetext}
where $\geta=\eta/(\epsilon+P)$ is the shear diffusion coefficient, $D_i=\sigma_{ii}/\chi_i$ are the vector/axial charge diffusion coefficients and the coefficient $\gsph=4\Gsph/(\chi_A T)$ describes dissipative effects due to sphaleron transitions. Since the right-hand side of Eq.~\eqref{eq:nA} contains a sum over all flavors, this contribution leads to an explicit coupling of different flavor components, which tends to erase the net axial charge in the system.

We first observe that Eqs. \eqref{eq:sound1} and \eqref{eq:sound2} are coupled and describe sound waves, whereas Eq. \eqref{eq:shear1} describes a purely diffusive shear mode. Eq. \eqref{eq:shear2} is also a diffusive shear mode, coupled to Eqs. \eqref{eq:nV} and \eqref{eq:nA}, which describe vector and axial charge density modes. We restrict our analysis to the coupled charge density equations, leaving out Eq. \eqref{eq:shear2} as the vector charge density fluctuations feed into the shear mode but the shear mode does not feed back into the charge density equations at linear order.

\subsection{Single flavor dynamics}
Before we address the more complex situation of multiple flavors, we will analyze the effect of sphaleron transitions on the coupled vector and axial charge dynamics of a single fermion flavor ($N_f=1$) with charge $q_{f}$ in the presence of a magnetic field.

\begin{figure*}
\includegraphics[width=0.45\linewidth]{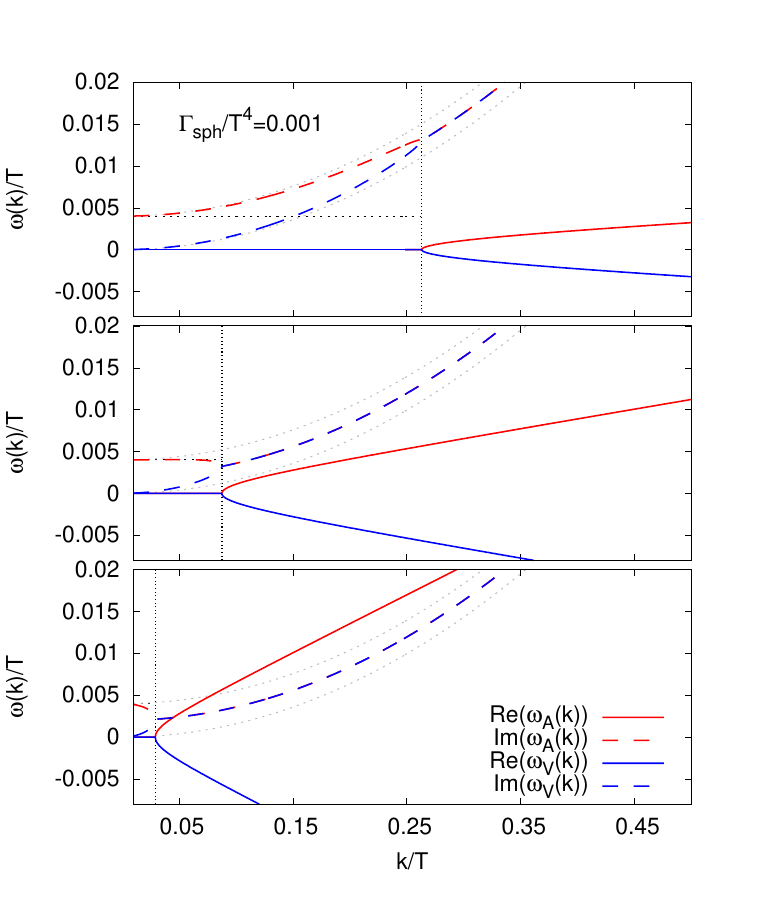}
\includegraphics[width=0.45\linewidth]{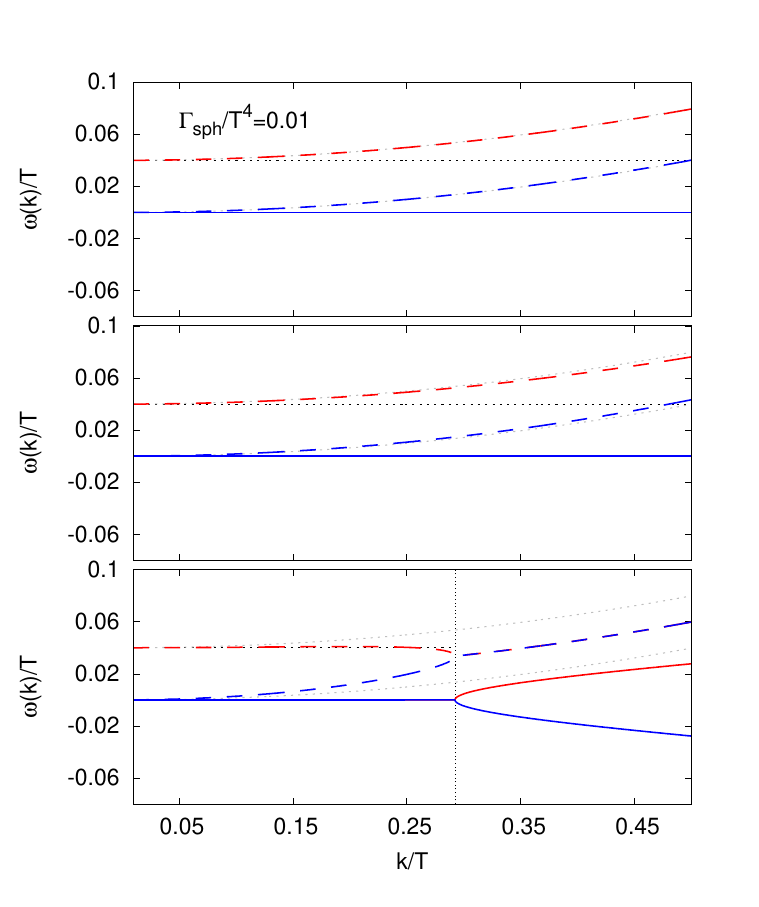}
\caption{Single-flavor dispersion relations $\omega_A$ \eqref{eq:dr_ax} and $\omega_V$ \eqref{eq:dr_vec} for different values of $\Gsph$ plotted for $\eb=0.05,0.15,0.45$ (top to bottom) respectively. Vertical dotted line represents $\kc$ for each case.}
\label{fig:sf_dispersions}
\end{figure*}

\begin{figure}
\includegraphics[width=\linewidth]{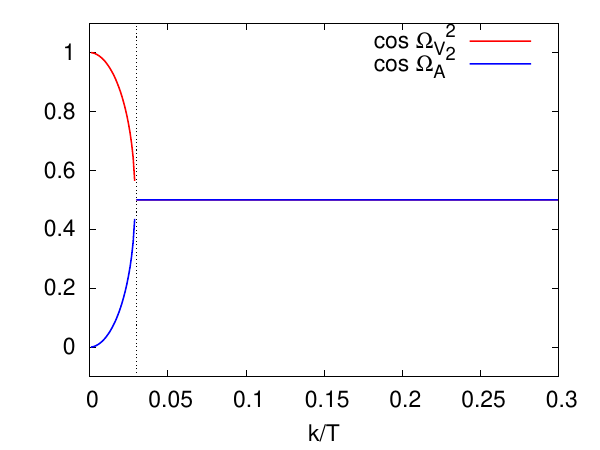}
\caption{Mixing angle $\cos\Omega_{A/V}^2$ for the single-flavor case plotted for $\eb=0.45$, $\Gsph/T^4=0.001$. Vertical dotted line represents $\kc$.}
\label{fig:sf_mixing}
\end{figure}

\begin{figure}
\includegraphics[width=\linewidth]{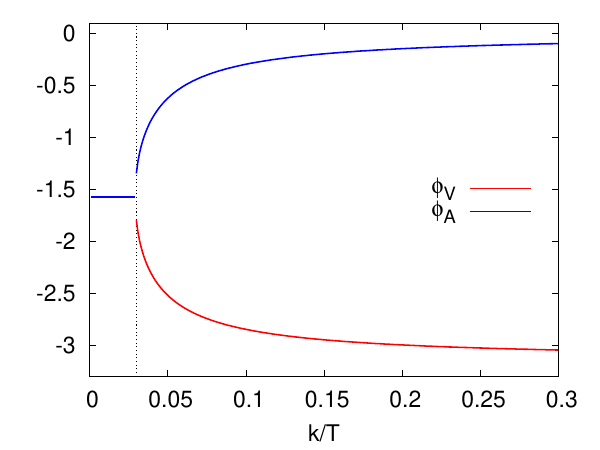}
\caption{Phases for the single-flavor case plotted for $\eb=0.45$, $\Gsph/T^4=0.001$. Vertical dotted line represents $\kc$.}
\label{fig:sf_phases}
\end{figure}

We rewrite Eqs. \eqref{eq:nV} and \eqref{eq:nA} in matrix form, using 
\begin{align}
M_{ab}^{N_{f}=1}=
\begin{pmatrix}
D\kt^2&ieq_fC\chi_A^{-1}\kt\cdot\Bt\\
ieq_fC\chi_V^{-1}\kt\cdot\Bt&D\kt^2+\gsph
\end{pmatrix} \label{eq:chargemat}
\end{align}
such that the fields $\phi_a=(\delta n_V,\delta n_A)$ satisfy the equation
\begin{align}
\partial_t\phi_a +M_{ab}\phi_b=0\;.
\label{eq:guvna}
\end{align}
By following standard procedure, the dispersion relations of the linearized hydrodynamic equations are then found by determining minus $i$ times the complex eigenvalues associated with the matrix  $M_{ab}^{N_{f}=1}$ in  Eq.~\eqref{eq:chargemat}.

Before studying the collective modes that emerge when explicitly accounting for the dissipative contribution of sphaleron transitions, we first address the dynamics of vector and axial charges in the absence of sphaleron transitions by setting $\gsph=0$ in Eq.~\eqref{eq:chargemat}. The resulting dispersion relations of the charge modes take the form
\begin{align}
\label{eq:CMWNoSph}
\omega_{\mp} &= -iD\kt^2\mp \frac{(eq_f)C}{\sqrt{\chi_A\chi_V}} |\kt\cdot\Bt|, 
\end{align}
which are the known dispersion relations associated with the CMW up to $\mathcal{O}(\kt^2)$ \cite{Kharzeev:2010gd}. We observe that the dispersion relations have two distinct, competing parts, namely a diffusive imaginary part and a propagating real part. Since the diffusion constant $D$ is fixed, the mechanism dominating the behavior of the excitations depends primarily on the magnitude and orientation of the wavevector ${\bf k}$ of the perturbation and on the strength of the magnetic field. In the presence of a weak magnetic field, the dynamics of charge modes will be governed by diffusion. As the magnetic field increases in strength, the low ${\bf k}$ modes oriented along the magnetic field will propagate with decreasing influence from diffusion.

We can further characterize the modes by discussing the associated eigenvectors,
\begin{align}
\vt_{\mp}=\begin{pmatrix}
\cos\Omega_{\mp}\\
e^{i\phi_{\mp}}\sin\Omega_{\mp}
\end{pmatrix},
\end{align}
in which the subscripted sign corresponds to the sign of the real part of the dispersion relations. The mixing angles $\Omega_{\mp}$ are  
\begin{align}
\tan\Omega_{\mp}=\sqrt{\frac{\chi_V}{\chi_A}},
\end{align}
and the phases $\phi_{\mp}$ are 
\begin{align}
e^{i\phi_-}= -1,\qquad e^{i\phi_+}= 1,
\end{align}
such that for equal vector/axial charge susceptibilities $\chi_{V} \approx \chi_{A}$, vector and axial evolution is maximally mixed.

When sphaleron transitions occur ($\gsph\neq0$), the dispersion relations can no longer be simply divided into a diffusive and a propagating part. Instead, the inclusion of sphaleron transitions associated with the term $\gsph$ leads to the emergence of a wavenumber threshold,
\begin{align}
\kc=\sqrt{\frac{\chi_V}{\chi_A}}\frac{2\Gsph}{e |q_f| C |\Bt|}, 
\label{eq:kC}
\end{align}
which provides the minimum wavenumber above which a propagating chiral magnetic wave (CMW) can form for a given magnetic field strength. Hence it is convenient to express the dispersion relations in terms of the characteristic scale $\kc$ as 
\begin{align}
\omega_{A}&= -\frac{i}{2}\left(\gsph+2D\kt^2\right)-\frac{\gsph}{2}\sqrt{\kkc^2-1} ,\quad \label{eq:dr_ax}\\
\omega_{V}&= -\frac{i}{2}\left(\gsph+2D\kt^2\right)+\frac{\gsph}{2}\sqrt{\kkc^2-1}, \label{eq:dr_vec}
\end{align}
where $\omega_A$ is the dispersion relation of the mode dominated by axial charge diffusion and $\omega_V$ is the dispersion relation of the mode dominated by vector charge diffusion. 

We plot the dispersion relations in Fig.~\ref{fig:sf_dispersions} for three different values of the magnetic field strength $\eb=0.05,0.15,0.45$ and further illustrate the behavior for two different values of the sphaleron transition rate, namely $\Gsph/T^4=0.001$, corresponding to perturbative estimates \cite{Moore:2010jd}, and $\Gsph/T^4=0.01$, which is on the order of recent (quneched) lattice QCD results~\cite{Altenkort:2020axj}. For illustrative purposes, we consider $N_c=3$ with the charge susceptibilities of the free theory, given by
\begin{align}
\chi_{A/V}(T,\mu_{A/V}=0)=N_c\frac{T^2}{3}=T^2,
\end{align}
in a charge-neutral plasma.

Each plot shows two distinct regimes separated by their respective value of $\kc$. Below $\kc$, CMWs cannot form and modes are purely dissipative as the dynamics is dominated by damping due to sphaleron transitions. On the other hand, above $\kc$, the dynamics of the modes depends on the magnetic field strength. As $\eb$ decreases, the wavenumber threshold for the formation of a CMW increases and dissipative effects increasingly dominate the propagation due to charge mixing in the presence of the magnetic field.
At high $\kc$, modes will form a CMW but is strongly damped due to the combined effects of sphaleron damping and charge diffusion. Only at sufficiently high $\eb$ can the CMW overcome the effects of sphaleron damping and propagate without significant dissipation, as seen in the lower left panel of Fig.~\ref{fig:sf_dispersions} for a small sphaleron transiton rate and large magnetic field strength. In the case of a large sphaleron rate, shown in the right panel of Fig.~\ref{fig:sf_dispersions} dissipative effects dominate for all magnetic field strength considered. Even for the larger magnetic field strength shown in the bottom right panel, the dominant effect of the vector/axial charge mixing is not the formation of a propagating CMW but rather the additional dissipative effects due to sphaleron transitions. 

We then investigate the extent of charge mixing by analyzing the corresponding eigenvectors, 
\begin{align}
\vt_{i}=\begin{pmatrix}
\cos\Omega_i\\
e^{i\phi_i}\sin\Omega_i
\end{pmatrix}.
\end{align}
for $i=A,V$. The mixing angles $\Omega_{A/V}$, shown in Fig. \ref{fig:sf_mixing},  characterize the mixing of vector and axial charged for the two modes. While at $k=0$ vector and axial charge dynamics is decoupled, a significant charge mixing already builds up in the dissipative regime $k<\kc$ regime, before for $k>\kc$, the mixing angle is identical for both modes, and the evolution of vector and axial charges is maximally mixed.

The phases $\phi_{A/V}$ are shown in Fig. \ref{fig:sf_phases}. In the regime where $k < \kc$, the phases are the same, $\phi_{V/A}= -\pi/2$. However, for $k>\kc$, as $k$ increases, the phases diverge towards a phase difference $\Delta \phi = \pi$. That is, $\phi_A$ approaches 0, whereas $\phi_V$ approaches $-\pi$.



\subsection{Multi flavor dynamics}

\begin{figure*}
\includegraphics[width=0.45\linewidth]{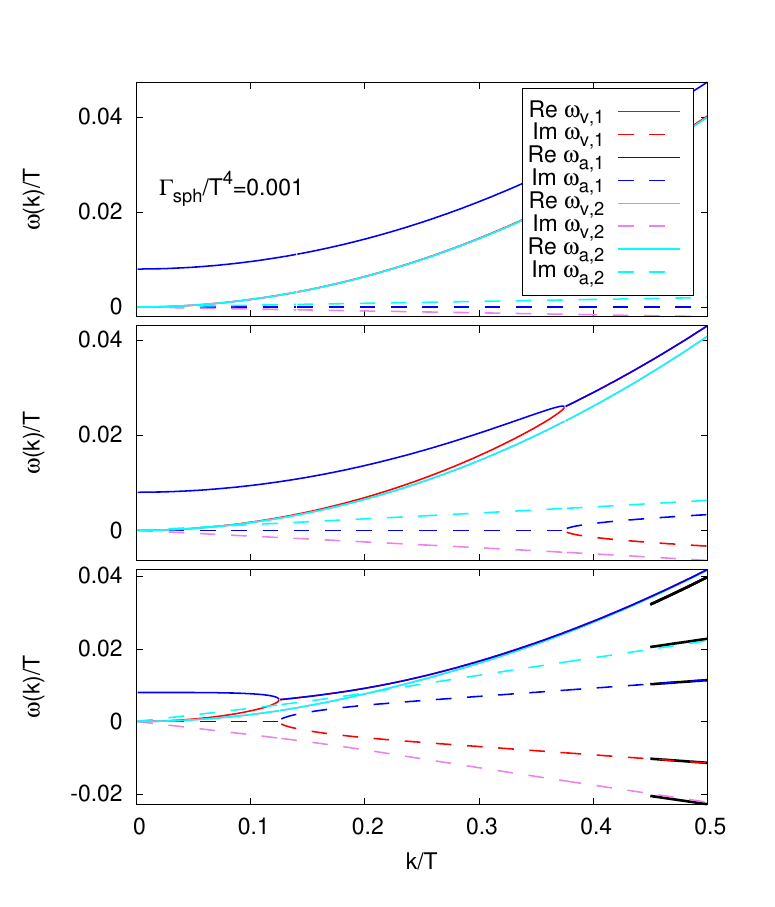}
\includegraphics[width=0.45\linewidth]{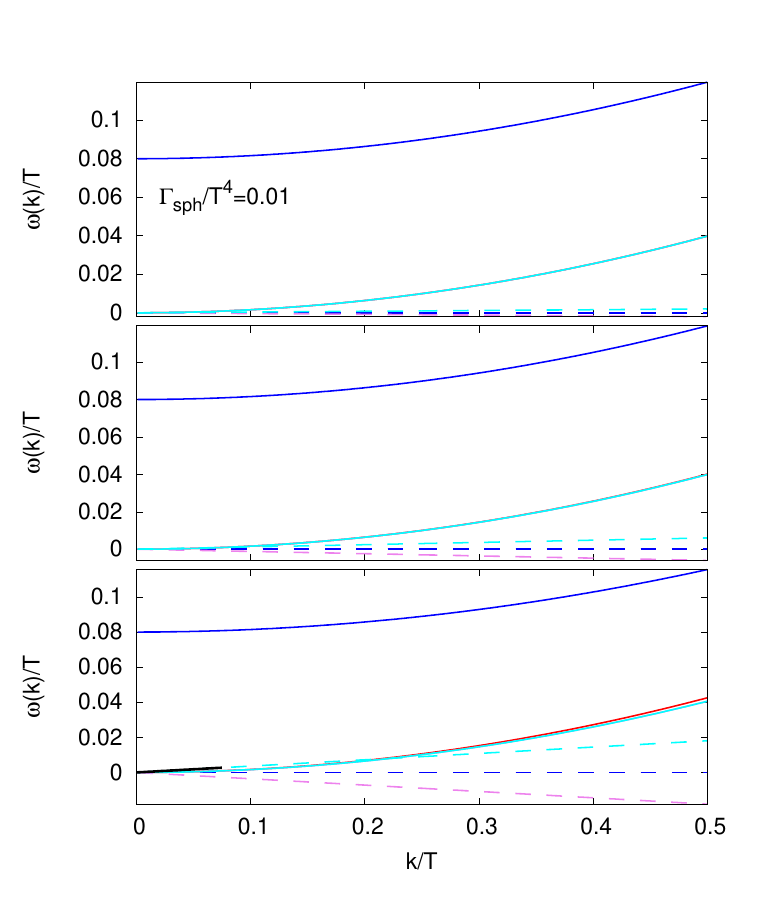}
\caption{Dispersion relations in the multiflavor case for different values of $\Gsph$ plotted for $\eb=0.05,0.15,0.45$ (top to bottom), respectively. Black solid lines indicate the asymptotic limits of small or large wavenumber $k$.}
\label{fig:mf_dispersions}
\end{figure*}

\begin{figure}
\includegraphics[width=\linewidth]{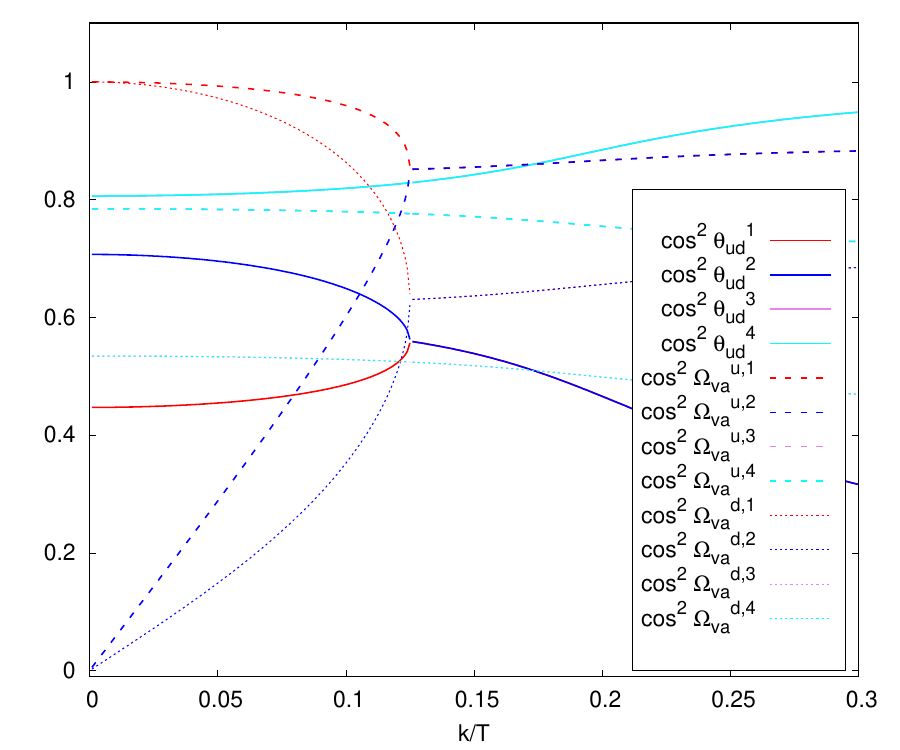}
\caption{Mixing angles for two-flavor system, $eB/T^2=0.45$.}
\label{fig:mf_mixing}
\end{figure}

\begin{figure}
\includegraphics[width=\linewidth]{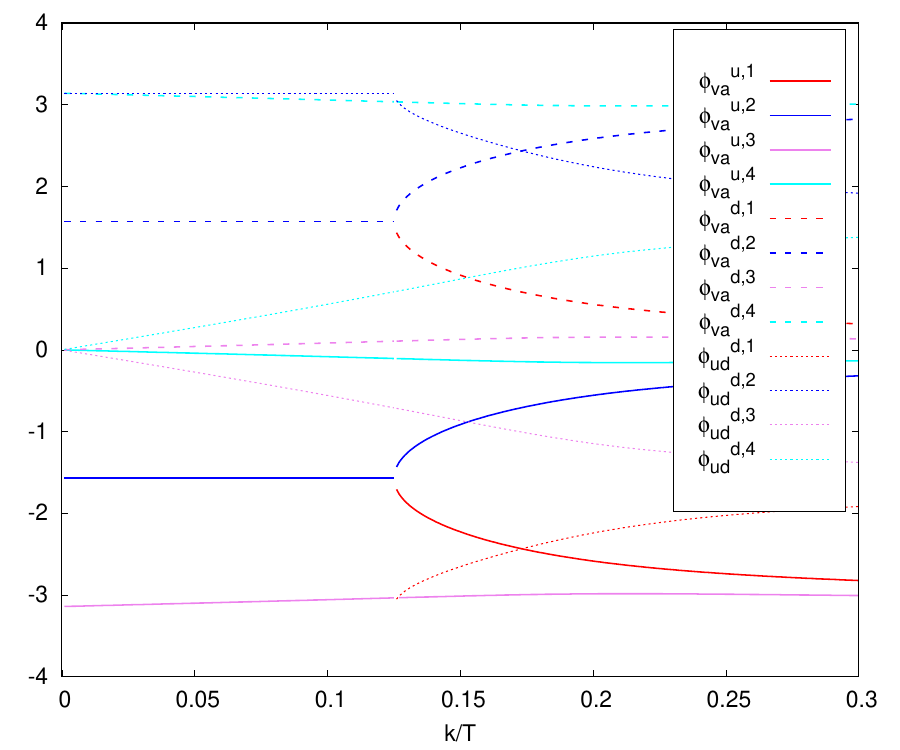}
\caption{Phases for two-flavor system, $eB/T^2=0.45$.}
\label{fig:mf_phases}
\end{figure}

We now move on and consider a two fermion-flavor system with up and down quarks. In this case, the evolution matrix for vector and axial charge dynamics of up and down quarks is given by
\begin{align}
M_{ab}^{N_f=2}=
\begin{pmatrix}
M_{ab}^{N_{f}=1}\big |_{q_f=q_u}& \begin{matrix}0&0\\0& \gsph \end{matrix}\\
\begin{matrix}0&0\\0& \gsph\end{matrix}& M_{ab}^{N_{f}=1}\big |_{q_f=q_d}
\end{pmatrix},
\label{eq:mf_matrix}
\end{align}
where $M_{ab}^{N_{f}=1}$ denotes the single-flavor matrix given in \eqref{eq:chargemat}, evaluated for the electric charge of the up quark and down quark, respectively. The dynamics is then governed by Eq. \eqref{eq:guvna} for $\phi_a=\{\delta n_{V,u},\delta n_{A,u},\delta n_{V,d},\delta n_{A,d}\}$, the vector and axial charge densities for up and down quarks. We emphasize that the dissipative term due to sphaleron transitions couples the dynamics of the up and down quarks, as can already be seen in Eq. \eqref{eq:nA}, where the right hand side is proportional to the net axial charge imbalance of all flavors.

The dispersion relations in the two flavor case are shown in Fig. \ref{fig:mf_dispersions} for both a low and higher sphaleron rate, for three different values of the magnetic field. As in the single-flavor case, we can express the eigenvectors in terms of mixing angles and phases. We parameterize the four eigenvectors via
\begin{align}
\vt_{i}=
\begin{pmatrix}
\cos\theta_{ud}\cos\Omega_{VA}^u\\
e^{i\varphi_{VA}^u}\cos\theta_{ud}\sin\Omega_{VA}^u\\
e^{i\varphi_{ud}}\sin\theta_{ud}\cos\Omega_{VA}^d \\
e^{i\varphi_{ud}}e^{i\varphi_{VA}^d}\sin\theta_{ud}\sin\Omega_{VA}^d
\end{pmatrix}, \quad i=\{1,2,3,4\}\;. \nonumber \\
\label{eq:mf_phases_param}
\end{align}
 Based on this parameterization, we find the mixing angles shown in Fig.~\ref{fig:mf_mixing} and the phases shown in Fig. \ref{fig:mf_phases}, where the mixing angle $\cos\theta_{ud}$ describes mixing between up and down flavors, while $\cos\Omega_{VA}^{u,d}$ describe axial and vector charge mixing. Since the general structure in Figs. \ref{fig:mf_dispersions},\ref{fig:mf_mixing} and \ref{fig:mf_phases} is rather complicated, we discuss the analytic forms of the vector/axial charge modes and dispersion relations in the two-flavor system  in the limiting cases of small and large wavenumber. 

In the large wavenumber limit, the sphaleron rate becomes increasingly less important, such that the asymptotic case is described when we take $\gsph \rightarrow 0$ in $M_{ab}^{N_f=2}$. In this case, there is no mixing between up and down flavors, such that the corresponding eigenvalues take the form
\begin{align}
\omega_{d\mp} &= -iD\kt^2\mp \frac{e|q_d|C}{\sqrt{\chi_A\chi_V}} |\kt\cdot\Bt|,\\
\omega_{u\mp} &= -iD\kt^2\mp \frac{e|q_u|C}{\sqrt{\chi_A\chi_V}} |\kt\cdot\Bt|,
\label{eq:highk_mf_evals}
\end{align}
which is identical to Eq.~(\ref{eq:CMWNoSph}) and describes the independent dynamics of up and down quarks. Conversely, in the small wavenumber limit ($k\rightarrow 0$) sphaleron damping plays a prominent role. In this limit, the leading eigenvalues are
\begin{align}
\omega_{1} = -2i\gsph, \quad \omega_2 = \omega_3 = \omega_4 = 0\;,
\label{eq:lowk_mf_evals}
\end{align} 
where the first mode corresponds to the relaxation of the net axial charge density ($\delta n_A^u+\delta n_A^d$) due to sphaleron transitions, while the axial charge difference between up and down quarks ($\delta n_A^u-\delta n_A^d$) is conserved, as well as the corresponding vector quantities, ($\delta n_V^u+\delta n_V^d$) and ($\delta n_V^u-\delta n_V^d$).
One can further disentangle the three degenerate eigenvalues by applying degenerate perturbation theory to next-to-leading order. By assuming $\chi_{a}=\chi_{V}=\chi$ for simplicty, and leaving details of the calculation for Appendix \ref{app:multiflavor}, the eigenvalues to first order in perturbation theory are given by
\begin{align}
\omega_2 = 0, \quad \omega_{3,4} = \mp \frac{eC}{\chi\sqrt{2}}\sqrt{(q_d^2+q_u^2)} |\kt\cdot\Bt|.\quad \ 
\end{align}
indicating the emergence of constant mode and conjugate pair of propagating chiral magnetic waves, which is indicated by a black line in the bottom right panel of Fig.~\ref{fig:mf_mixing}. 

Generally beyond these two simple limits, the coupled dynamics is rather complicated, as can be inferred from the rather complex structures seen in Figs. (\ref{fig:mf_dispersions}), (\ref{fig:mf_mixing}), and (\ref{fig:mf_phases}). Clearly, the reason for this is that, in the multi-flavor case, even in the the small $k$ limit, all of the modes are associated with linear combinations of $u$ and $d$ vector and axial charges, as can be deduced the analytic expressions for the eigenvectors in Appendix \ref{app:multiflavor} as well as from the mixing angles and phases in Figs.~\ref{fig:mf_mixing} and \ref{fig:mf_phases}.

\section{Effects of sphaleron damping on vector and axial charge dynamics}\label{sec:diffusion} 

\begin{figure*}
\centering
\subfloat{\includegraphics[width=\textwidth]{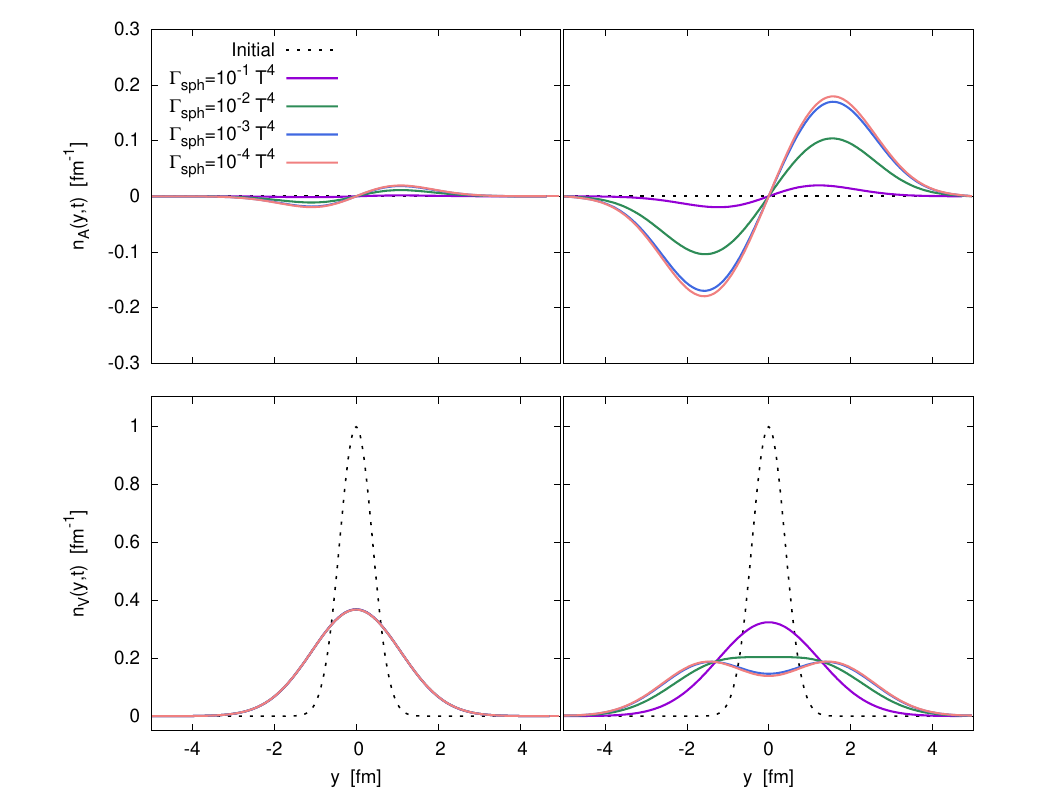}}
\caption{Spatial profiles of axial ($n_{A}$) and vector ($n_{V}$) charge distributions for \emph{initial vector charge perturbation} after an evolution for $t=10$ fm/c. Different curves in each panel correspond to four different values of the sphaleron transition rate $\Gamma_{\rm sph}$. Different columns show the the results for different magnetic field strength $\eb=1/16$ in the left column and $\eb=1$ in the right column.}
\label{fig:init-nV}
\end{figure*}

\begin{figure*}
\centering
\subfloat{\includegraphics[width=\textwidth]{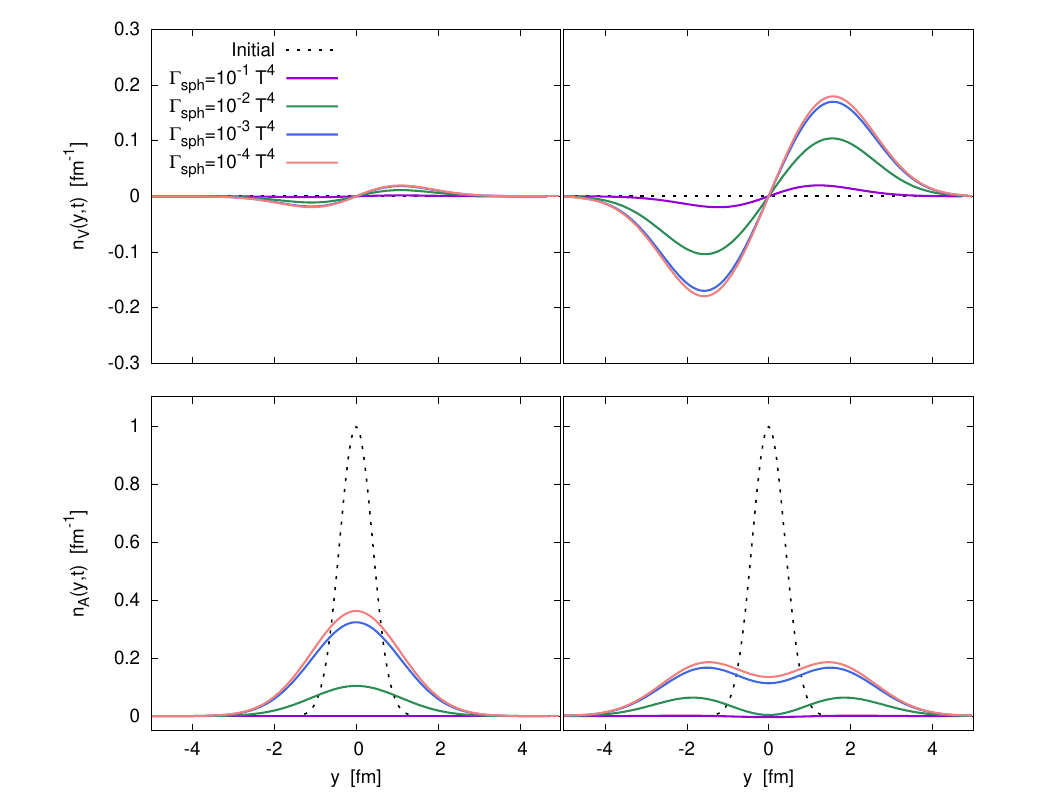}}
\caption{Spatial profiles of axial ($n_{A}$) and vector ($n_{V}$) charge distributions for \emph{initial axial charge perturbation} after an evolution for $t=10$ fm/c. Different curves in each panel correspond to four different values of the sphaleron transition rate $\Gamma_{\rm sph}$. Different columns show the the results for different magnetic field strength $\eb=1/16$ in the left column and $\eb=1$ in the right column.}
\label{fig:init-nA}
\end{figure*}


\begin{figure}
\includegraphics[width=\linewidth]{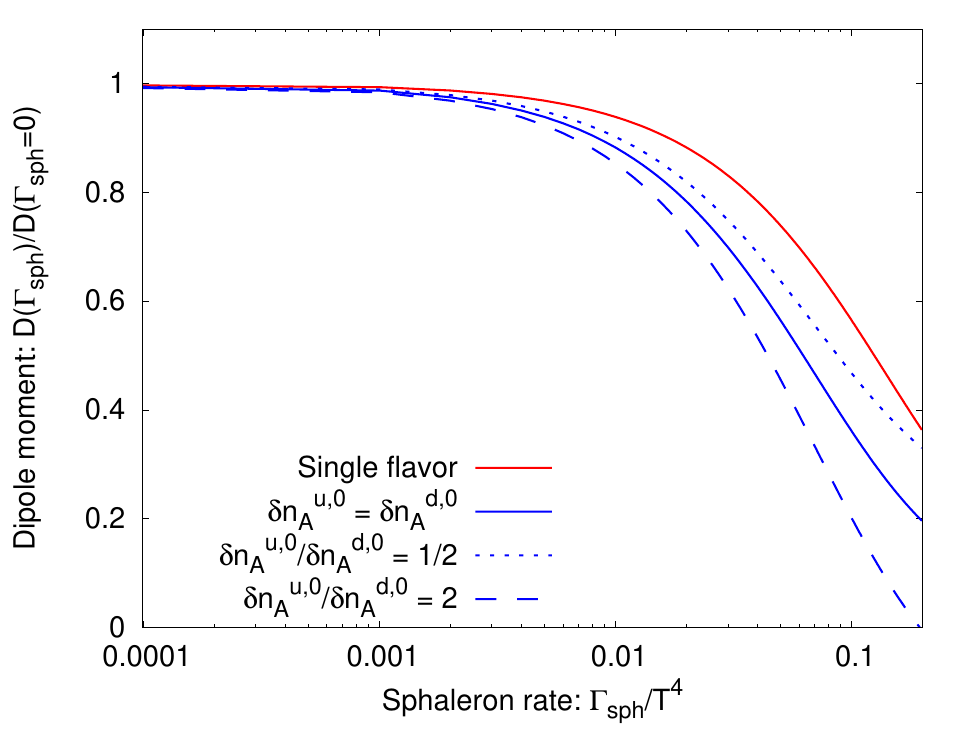}
\caption{Electric charge separation, quantified by the electric dipole moment $D$ for an initial axial charge distribution as a function of $\Gsph$ for single- and two-flavor configurations of various initial charge ratios at $t=10\ {\rm fm}/c$. }
\label{fig:dipole2}
\end{figure}

Next, in order to assess the impact of sphaleron transitions on normal and anomalous transport phenomena in a QCD plasma, we investigate the response of the system to an initial charge inhomogeneity by solving the linearized hydrodynamic equations \eqref{eq:guvna} numerically. We orient the magnetic field along the $y$-direction and study perturbations in the $x-y$ plane to loosely mimic the evolution in the transverse plane in an off-central heavy-ion collision.\footnote{When solving Eq.~\eqref{eq:guvna} numerically, in practice we discretize the evolution on a two-dimensional spatial lattice ($256^2$).  The lattice is scaled such that the length of the sides were $10$~fm with spacing $a_S=10/256$~fm.} We set the scale by setting temperature $T=4T_C$, where $T_C=155$ MeV is the QCD cross-over temperature and study the evolution over a time scale $t=10$ fm/c. We limit ourselves to the single-flavor scenario ($N_f=1$) and consider two magnetic field strength regimes: $\eb=1/16$ and $\eb=1$. The first of these regimes, where $\eb=1/16$, was chosen to correspond to $m_{\pi}^2$, an optimistic estimate for the magnetic field strength achieved in a heavy ion collision~\cite{Skokov:2009qp}. The second, $\eb=1$, was chosen arbitrarily such that it was much stronger than $m_{\pi}^2$. We consider four different values of the sphaleron rate $\Gsph$ for each magnetic field strength $\eb$,  and monitor the evolution of the vector/axial charge distributions along the magnetic field direction, i.e. $n_{V/A}(y,t)=\int_{x,z} n_{V/A}(x,y,z,t)$, to probe how sphaleron transitions affect vector and axial charge transport.

\subsection{Vector charge perturbations}
We first consider an initial vector charge perturbation, given by a Gaussian distribution of width $\sigma=0.4R_p$, $R_{p}=1$ fm, such that the width is on the order of the size of a nucleon -- the characteristic length scale of variations in the transverse plane of a heavy-ion collision. By studying the vector and axial charge profiles after $t=10$ fm/c of evolution as depicted in Fig.~\ref{fig:init-nV}, we observe that vector charge diffuses while axial charge separates along the direction of the magnetic field. At $\eb=1/16$, the vector charge diffuses with no discernible difference with respect to the value of the sphaleron transition rate. However, when the magnetic field strength is increased to $\eb=1$, the charge either purely diffuses or forms a highly diffusive wave. This behavior depends on the sphaleron transition rate.  In fact, there is a clear transition in behavior between the charge distribution for $\Gsph/T^4=10^{-3}$ and $\Gsph/T^4=10^{-2}$; as the sphaleron rate increases, the behavior of the vector charge changes from diffusive propagation to purely diffusive. On the other hand, axial charge separates in the same manner for both magnetic field strengths, though the magnitude of charge separation is greater for a strong magnetic field. One sees immediately from Fig. \ref{fig:init-nV} that the magnitude and distance of charge separation depend on the value of the sphaleron transition rate. Hence, as the rate of sphaleron transitions increases, the magnitude and distance of separation decreases.

\subsection{Axial charge perturbations}
Next we consider an initial perturbation of the axial charge density, which can be seen as a simple toy model for dynamics of the Chiral Magnetic Effect (CME) in heavy ion collisions~\cite{Fukushima:2008xe}. We employ the same parameters as for the initial vector charge perturbation, and present our results for the vector and axial charge profiles in Fig.~\ref{fig:init-nA}. As can be expected, the response to an initial axial charge perturbation is significantly more sensitive to the sphaleron rate. Specifically, for $\eb=1/16$ shown in the left column, the axial charge profile diffuses and decays and the decay rate depends on the sphaleron rate. For $\eb=1$, the modes for each of the four sphaleron rates form a decaying and highly diffusive wave. With regards to vector charge for the initial axial distribution, one clearly observes a separation of vector charges along the direction of the magnetic field, albeit the amount of charge separation strongly depends on the sphaleron rate and the magnetic field strength. 

We also find that for the weaker field case, $\eb=1/16$ where the magnetic field strength is relevant for heavy ion collisions, the axial charge charge only diffuses and there is no clear sign of propagating waves. Even though a small amount of vector charge separation is still generated also in this case, it is clear that dissipative effects dominate in this case, and clearly need to be taken into account in a realistic description of the dynamcis of CME and CMW in heavy-ion collisions.

\subsection{Sensitivity of charge separation to the sphaleron rate} Importantly, the vector charge separation along the direction of the magnetic field has been suggested as an experimental signature of the CME in heavy-ion collisions~\cite{Koch:2016pzl}. Since this   charge separation is sensitive to the sphaleron transition rate, we will further quantify this dependence by using the dipole moment. Specifically, we consider vector charge separation as the result of an initial axial charge perturbation, and determine the electric dipole moment 
\begin{align}
D(\Bt,t)&=\int d^3x\ \frac{\xt\cdot\Bt}{|\Bt|} \sum_{f} eq_{f} n_{V,f}(t,\xt),
\label{eq:dipole-form}
\end{align}
which quantifies the amount of electric charge separation along the direction of the magnetic field.

We first derive an expression for the dipole moment for the case of a single quark flavor ($N_f=1$), rewriting Eq. (\ref{eq:dipole-form}) as
\begin{align}
D(\Bt,t)&=\int d^3x\ \frac{\xt\cdot\Bt}{|\Bt|}\int \frac{d^3k}{(2\pi)^3} e^{i\kt\cdot\xt} 
\begin{pmatrix}
1 \\
0
\end{pmatrix}^t \nonumber \\
&\times \exp\left\{M_{ab}(\kt)(t-t_0)\right\} 
\begin{pmatrix}
\Tilde{n}_V(t_0,\kt) \\
\Tilde{n}_A(t_0,\kt) 
\end{pmatrix},
\end{align}
where we have used the notation $\Tilde{n}_{V/A}(\kt)$ to differentiate between the charge distributions in coordinate space and their Fourier-transformed counterparts. Recall, matrix $M_{ab}$ is defined by Eq. \eqref{eq:chargemat}. Then, switching the order of integration,
\begin{align}
D(\Bt,t)&=&\int \frac{d^3k}{(2\pi)^3}\int d^3x \frac{\left(-i\partial_{\kt}^je^{i\kt\cdot\xt}\right)B_j}{|\Bt|}
\begin{pmatrix}
1 \\
0
\end{pmatrix}^t \nonumber \\
&&\times \exp\left\{M_{ab}(\kt)(t-t_0)\right\} 
\begin{pmatrix}
\Tilde{n}_V(t_0,\kt) \\
\Tilde{n}_A(t_0,\kt) 
\end{pmatrix},
\end{align}
which becomes 
\begin{align}
D(\Bt,t)=&\left(\int \frac{d^3k}{(2\pi)^3}\int d^3x e^{i\kt\cdot\xt}\right)\frac{i\partial_{\kt}^jB_j}{|\Bt|}\begin{pmatrix}
1 \\
0
\end{pmatrix}^t \nonumber \\
&\times \exp\left\{M_{ab}(\kt)(t-t_0)\right\} 
\begin{pmatrix}
\Tilde{n}_V(t_0,\kt) \\
\Tilde{n}_A(t_0,\kt) 
\end{pmatrix}\nonumber \\
&= \frac{i\partial_{\kt}^jB_j}{|\Bt|}\begin{pmatrix}
1 \\
0
\end{pmatrix}^t \exp\left\{M_{ab}(\kt)(t-t_0)\right\} \nonumber\\
&\times\begin{pmatrix}
\Tilde{n}_V(t_0,\kt) \\
\Tilde{n}_A(t_0,\kt) 
\end{pmatrix}\biggr|_{k=0}.\label{eq:dipole-eval}
\end{align}
Evaluating Eq. \eqref{eq:dipole-eval} for an initial axial charge perturbation and one quark flavor ($N_f=1$), we find the dipole moment
\begin{align}
D(\Bt,t)&=\frac{-eq_{f}|\Bt|CT}{4\Gsph} \left(1-e^{-4\Gsph t/(\chi_A T)}\right)\delta \tilde{n}_A^{0},\nonumber\\
\label{eq:sf-dipole}
\end{align}
where we use $\delta \tilde{n}_A^{0}=\delta \tilde{n}_A(t=0,\kt=0)$ to denote the initial net axial charge imbalance.

Similarly, in the case of two quark flavors ($N_f=2$), we find
\begin{align}
\label{eq:mf-dipole}
&D(\Bt,t)= \\ & \frac{-e|\Bt|C(q_u^2+q_d^2)T}{16\Gsph}\Big(1-e^{-\frac{8\Gsph t}{\chi_A T}} \Big)
(\delta \tilde{n}_A^{u,0}+\delta \tilde{n}_A^{d,0}) \nonumber \\
&-
\frac{e|\Bt|C(q_u^2-q_d^2)T}{2\Gsph}
\Big(\frac{\Gsph t}{\chi_A T}e^{-\frac{8\Gsph t}{\chi_A T}}\Big)(\delta \tilde{n}_A^{u,0}-\delta \tilde{n}_A^{d,0}), \nonumber
\end{align}
where as before $\delta \tilde{n}_A^{u/d,0}=\delta \tilde{n}_A^{u/d}(t=0,\kt=0)$ denote the initial axial charge imbalance of $u$ and $d$ flavors, such that the terms in the second line describe the respone to a net axial charge imbalance of both flavors, whereas the terms in the third line describe the response to an axial charge difference between $u$ and $d$ flavors. However, most importantly, from Eqs. \eqref{eq:sf-dipole} and \eqref{eq:mf-dipole}, we immediately see the relationship between the sphaleron transition rate $\Gsph$ and separation of charge, as quantified by the dipole moment $D(\Bt,t)$. 

We illustrate the relations for both the single-flavor and two-flavor case in Fig.~\ref{fig:dipole2}, where we present the dependence of the dipole moment $D({\bf B},t)$ on the sphaleron transition rate $\Gsph$. By normalizing the dipole moment to its value for $\Gsph=0$, the quantity $D(\Gsph)/D(\Gsph=0)$ becomes independent of the magnetic field strength (c.f. Eqns.~(\ref{eq:sf-dipole},\ref{eq:mf-dipole})) and can be viewed an overall suppression factor of the charge separation signal due to sphaleron transitions. When the sphaleron transition rate is large, all terms proportional to $e^{-\#\Gsph t/\chi_{a}T}$ in Eqns.~(\ref{eq:sf-dipole},\ref{eq:mf-dipole}) can be dropped and the charge separation is proportional to $1/\Gsph$.  By inspecting the results in Fig.~\ref{fig:dipole2} one finds that after an evolution for $10 {\rm fm}/c$, the suppression for sphaleron rates $\Gsph/T^4\lesssim 0.01$ is still rather modest. However, for values on the order of the (quenched) lattice QCD estimates~\cite{Altenkort:2020axj}  $\Gsph/T^4\gtrsim 0.02$ there is in a significant suppression of the signal, as well as a strong sensitivity of the result to the actual value of the sphaleron transition rate. While such a suppression may make it harder to detect possible signatures of the CME and CMW in heavy-ion collisions, the strong sensitivity to the sphaleron rate also suggests a possible experimental avenue for constraining the sphaleron rate using charge separation measurements associated with chiral phenomena such as the CME and CMW. 

\section{Conclusions \& Outlook}\label{sec:conclusion}
Based on a general discussion of the criteria for the validity of a macroscopic description of the axial charge dynamics in high-temperature QCD plasmas, we modified the anomalous hydrodynamic equations of motion to explicitly include dissipative effects sourced by sphaleron transitions. Within this framework, dissipation due to sphaleron transitions is incorporated as a damping term proportional to the sphaleron transition rate, which depletes the net axial charge imbalance of all fermion flavors and contributes positively to entropy production in the system. Noteably, in the case of multi-flavors the dissipative contribution from sphaleron damping also coupled the dynamics of different flavors, as the dissipative term is proportional to the sum of the axial charge density of all flavors.  

By linearizing the hydrodynamic equations around a space-time independent background, we investigated the coupled dynamics of vector and axial charge perturbations in a charge neutral background and contrasted our results including sphaleron damping to the traditional behavior of a Chiral Magnetic Wave. When sphaleron damping is taken into consideration, a characteristic wavenumber scale $\kc$ emerges. Below $\kc$, charge modes experience decaying diffusive behavior as the dynamics is dominated by sphaleron transitions. Conversely, above $\kc$, the modes behave like an ordinary CMW, albeit the latter is typically strongly damped. The threshold $\kc \sim \Gamma_{\rm sph}/eB$ depends on the sphaleron transition rate and magnetic field strength and we expect that for typical values achieved in heavy-ion collisions at RHIC and LHC energies dissipative effects dominate and the decaying diffusive behavior is realized.

By studying the time evolution of linearized vector/axial charge perturbations, we visualized the impact of the sphaleron transition rate on vector and axial charge transport in a QCD plasma in the presence of a magnetic field. Strikingly, for sufficiently strong magnetic fields, the sphaleron transition rate also has an impact on vector charge transport, i.e. the vector charge response to a vector charge perturbation, which may be interesting from the point of view of extracting the QCD sphaleron rate on the lattice. Secondly, in the presence of an axial charge imbalance, we observe the expected separation of vector charges along the direction of the magnetic field. Even though the amount of charge separation  strongly depends on the sphaleron rate and magnetic field strength, the general phenomenon of charge separation persists. We further quantified the amount of charge separation in terms of the electric dipole moment, and determined its dependence on the sphaleron rate. We find that for realistic values of the sphaleron transition rate, the charge separation can easily be suppressed by a factor of two compared to the situation where dissipative effects due to sphaleron transitions are not taken into account.

Since the charge separation is highly sensitive to the sphaleron transition rate, it is conceivable that experimental measurements of charge separation can be used to constrain the QCD sphaleron rate. Such constraints would not only be useful to confront current state-of-the-art calculations, but would provide a unique measurements that can elucidate topological properties of QCD. Our results thus motivate the development of a more comprehensive treatment of axial charge dynamics, where it would also be important to extend the present framework to include fluctuations of axial charge sourced by sphaleron transitions.


\textit{Acknowledgements: }  We thank G.~D.~Moore, D.E.~Kharveez and S.~Sharma for insightful discussions.
LdB is supported by the Deutsche Forschungsgemeinschaft (DFG, German Research Foundation) through the Collaborative Research Center, Project-ID 27381115, SFB 1225 ISOQUANT. SS is supported by the Deutsche Forschungsgemeinschaft (DFG, German Research Foundation) through the CRC-TR 211 ‘Strong-interaction matter under extreme conditions’ - project number:
315477589 TRR-211.

\bibliographystyle{apsrev4-1}
\bibliography{main}

\begin{thebibliography}{31}%
\makeatletter
\providecommand \@ifxundefined [1]{%
 \@ifx{#1\undefined}
}%
\providecommand \@ifnum [1]{%
 \ifnum #1\expandafter \@firstoftwo
 \else \expandafter \@secondoftwo
 \fi
}%
\providecommand \@ifx [1]{%
 \ifx #1\expandafter \@firstoftwo
 \else \expandafter \@secondoftwo
 \fi
}%
\providecommand \natexlab [1]{#1}%
\providecommand \enquote  [1]{``#1''}%
\providecommand \bibnamefont  [1]{#1}%
\providecommand \bibfnamefont [1]{#1}%
\providecommand \citenamefont [1]{#1}%
\providecommand \href@noop [0]{\@secondoftwo}%
\providecommand \href [0]{\begingroup \@sanitize@url \@href}%
\providecommand \@href[1]{\@@startlink{#1}\@@href}%
\providecommand \@@href[1]{\endgroup#1\@@endlink}%
\providecommand \@sanitize@url [0]{\catcode `\\12\catcode `\$12\catcode
  `\&12\catcode `\#12\catcode `\^12\catcode `\_12\catcode `\%12\relax}%
\providecommand \@@startlink[1]{}%
\providecommand \@@endlink[0]{}%
\providecommand \url  [0]{\begingroup\@sanitize@url \@url }%
\providecommand \@url [1]{\endgroup\@href {#1}{\urlprefix }}%
\providecommand \urlprefix  [0]{URL }%
\providecommand \Eprint [0]{\href }%
\providecommand \doibase [0]{http://dx.doi.org/}%
\providecommand \selectlanguage [0]{\@gobble}%
\providecommand \bibinfo  [0]{\@secondoftwo}%
\providecommand \bibfield  [0]{\@secondoftwo}%
\providecommand \translation [1]{[#1]}%
\providecommand \BibitemOpen [0]{}%
\providecommand \bibitemStop [0]{}%
\providecommand \bibitemNoStop [0]{.\EOS\space}%
\providecommand \EOS [0]{\spacefactor3000\relax}%
\providecommand \BibitemShut  [1]{\csname bibitem#1\endcsname}%
\let\auto@bib@innerbib\@empty
\bibitem [{\citenamefont {Brandenburg}\ \emph {et~al.}(2017)\citenamefont
  {Brandenburg}, \citenamefont {Schober}, \citenamefont {Rogachevskii},
  \citenamefont {Kahniashvili}, \citenamefont {Boyarsky}, \citenamefont
  {Frohlich}, \citenamefont {Ruchayskiy},\ and\ \citenamefont
  {Kleeorin}}]{Brandenburg:2017rcb}%
  \BibitemOpen
  \bibfield  {author} {\bibinfo {author} {\bibfnamefont {A.}~\bibnamefont
  {Brandenburg}}, \bibinfo {author} {\bibfnamefont {J.}~\bibnamefont
  {Schober}}, \bibinfo {author} {\bibfnamefont {I.}~\bibnamefont
  {Rogachevskii}}, \bibinfo {author} {\bibfnamefont {T.}~\bibnamefont
  {Kahniashvili}}, \bibinfo {author} {\bibfnamefont {A.}~\bibnamefont
  {Boyarsky}}, \bibinfo {author} {\bibfnamefont {J.}~\bibnamefont {Frohlich}},
  \bibinfo {author} {\bibfnamefont {O.}~\bibnamefont {Ruchayskiy}}, \ and\
  \bibinfo {author} {\bibfnamefont {N.}~\bibnamefont {Kleeorin}},\ }\href
  {\doibase 10.3847/2041-8213/aa855d} {\bibfield  {journal} {\bibinfo
  {journal} {Astrophys. J. Lett.}\ }\textbf {\bibinfo {volume} {845}},\
  \bibinfo {pages} {L21} (\bibinfo {year} {2017})},\ \Eprint
  {http://arxiv.org/abs/1707.03385} {arXiv:1707.03385 [astro-ph.CO]}
  \BibitemShut {NoStop}%
\bibitem [{\citenamefont {Koch}\ \emph
  {et~al.}(2017{\natexlab{a}})\citenamefont {Koch}, \citenamefont
  {Schlichting}, \citenamefont {Skokov}, \citenamefont {Sorensen},
  \citenamefont {Thomas}, \citenamefont {Voloshin}, \citenamefont {Wang},\ and\
  \citenamefont {Yee}}]{Skokov:2016yrj}%
  \BibitemOpen
  \bibfield  {author} {\bibinfo {author} {\bibfnamefont {V.}~\bibnamefont
  {Koch}}, \bibinfo {author} {\bibfnamefont {S.}~\bibnamefont {Schlichting}},
  \bibinfo {author} {\bibfnamefont {V.}~\bibnamefont {Skokov}}, \bibinfo
  {author} {\bibfnamefont {P.}~\bibnamefont {Sorensen}}, \bibinfo {author}
  {\bibfnamefont {J.}~\bibnamefont {Thomas}}, \bibinfo {author} {\bibfnamefont
  {S.}~\bibnamefont {Voloshin}}, \bibinfo {author} {\bibfnamefont
  {G.}~\bibnamefont {Wang}}, \ and\ \bibinfo {author} {\bibfnamefont {H.-U.}\
  \bibnamefont {Yee}},\ }\href {\doibase 10.1088/1674-1137/41/7/072001}
  {\bibfield  {journal} {\bibinfo  {journal} {Chin. Phys. C}\ }\textbf
  {\bibinfo {volume} {41}},\ \bibinfo {pages} {072001} (\bibinfo {year}
  {2017}{\natexlab{a}})},\ \Eprint {http://arxiv.org/abs/1608.00982}
  {arXiv:1608.00982 [nucl-th]} \BibitemShut {NoStop}%
\bibitem [{\citenamefont {Li}\ \emph {et~al.}(2016)\citenamefont {Li},
  \citenamefont {Kharzeev}, \citenamefont {Zhang}, \citenamefont {Huang},
  \citenamefont {Pletikosic}, \citenamefont {Fedorov}, \citenamefont {Zhong},
  \citenamefont {Schneeloch}, \citenamefont {Gu},\ and\ \citenamefont
  {Valla}}]{Li:2014bha}%
  \BibitemOpen
  \bibfield  {author} {\bibinfo {author} {\bibfnamefont {Q.}~\bibnamefont
  {Li}}, \bibinfo {author} {\bibfnamefont {D.~E.}\ \bibnamefont {Kharzeev}},
  \bibinfo {author} {\bibfnamefont {C.}~\bibnamefont {Zhang}}, \bibinfo
  {author} {\bibfnamefont {Y.}~\bibnamefont {Huang}}, \bibinfo {author}
  {\bibfnamefont {I.}~\bibnamefont {Pletikosic}}, \bibinfo {author}
  {\bibfnamefont {A.~V.}\ \bibnamefont {Fedorov}}, \bibinfo {author}
  {\bibfnamefont {R.~D.}\ \bibnamefont {Zhong}}, \bibinfo {author}
  {\bibfnamefont {J.~A.}\ \bibnamefont {Schneeloch}}, \bibinfo {author}
  {\bibfnamefont {G.~D.}\ \bibnamefont {Gu}}, \ and\ \bibinfo {author}
  {\bibfnamefont {T.}~\bibnamefont {Valla}},\ }\href {\doibase
  10.1038/nphys3648} {\bibfield  {journal} {\bibinfo  {journal} {Nature Phys.}\
  }\textbf {\bibinfo {volume} {12}},\ \bibinfo {pages} {550} (\bibinfo {year}
  {2016})},\ \Eprint {http://arxiv.org/abs/1412.6543} {arXiv:1412.6543
  [cond-mat.str-el]} \BibitemShut {NoStop}%
\bibitem [{\citenamefont {Bell}\ and\ \citenamefont
  {Jackiw}(1969)}]{Bell:1969ts}%
  \BibitemOpen
  \bibfield  {author} {\bibinfo {author} {\bibfnamefont {J.~S.}\ \bibnamefont
  {Bell}}\ and\ \bibinfo {author} {\bibfnamefont {R.}~\bibnamefont {Jackiw}},\
  }\href {\doibase 10.1007/BF02823296} {\bibfield  {journal} {\bibinfo
  {journal} {Nuovo Cim. A}\ }\textbf {\bibinfo {volume} {60}},\ \bibinfo
  {pages} {47} (\bibinfo {year} {1969})}\BibitemShut {NoStop}%
\bibitem [{\citenamefont {Adler}(1969)}]{Adler:1969gk}%
  \BibitemOpen
  \bibfield  {author} {\bibinfo {author} {\bibfnamefont {S.~L.}\ \bibnamefont
  {Adler}},\ }\href {\doibase 10.1103/PhysRev.177.2426} {\bibfield  {journal}
  {\bibinfo  {journal} {Phys. Rev.}\ }\textbf {\bibinfo {volume} {177}},\
  \bibinfo {pages} {2426} (\bibinfo {year} {1969})}\BibitemShut {NoStop}%
\bibitem [{\citenamefont {Horvath}\ \emph {et~al.}(2020)\citenamefont
  {Horvath}, \citenamefont {Hou}, \citenamefont {Liao},\ and\ \citenamefont
  {Ren}}]{Horvath:2019dvl}%
  \BibitemOpen
  \bibfield  {author} {\bibinfo {author} {\bibfnamefont {M.}~\bibnamefont
  {Horvath}}, \bibinfo {author} {\bibfnamefont {D.}~\bibnamefont {Hou}},
  \bibinfo {author} {\bibfnamefont {J.}~\bibnamefont {Liao}}, \ and\ \bibinfo
  {author} {\bibfnamefont {H.-c.}\ \bibnamefont {Ren}},\ }\href {\doibase
  10.1103/PhysRevD.101.076026} {\bibfield  {journal} {\bibinfo  {journal}
  {Phys. Rev. D}\ }\textbf {\bibinfo {volume} {101}},\ \bibinfo {pages}
  {076026} (\bibinfo {year} {2020})},\ \Eprint
  {http://arxiv.org/abs/1911.00933} {arXiv:1911.00933 [hep-ph]} \BibitemShut
  {NoStop}%
\bibitem [{\citenamefont {Stephanov}\ \emph {et~al.}(2015)\citenamefont
  {Stephanov}, \citenamefont {Yee},\ and\ \citenamefont
  {Yin}}]{Stephanov:2014dma}%
  \BibitemOpen
  \bibfield  {author} {\bibinfo {author} {\bibfnamefont {M.}~\bibnamefont
  {Stephanov}}, \bibinfo {author} {\bibfnamefont {H.-U.}\ \bibnamefont {Yee}},
  \ and\ \bibinfo {author} {\bibfnamefont {Y.}~\bibnamefont {Yin}},\ }\href
  {\doibase 10.1103/PhysRevD.91.125014} {\bibfield  {journal} {\bibinfo
  {journal} {Phys. Rev. D}\ }\textbf {\bibinfo {volume} {91}},\ \bibinfo
  {pages} {125014} (\bibinfo {year} {2015})},\ \Eprint
  {http://arxiv.org/abs/1501.00222} {arXiv:1501.00222 [hep-th]} \BibitemShut
  {NoStop}%
\bibitem [{\citenamefont {Jimenez-Alba}\ \emph {et~al.}(2014)\citenamefont
  {Jimenez-Alba}, \citenamefont {Landsteiner},\ and\ \citenamefont
  {Melgar}}]{Jimenez-Alba:2014iia}%
  \BibitemOpen
  \bibfield  {author} {\bibinfo {author} {\bibfnamefont {A.}~\bibnamefont
  {Jimenez-Alba}}, \bibinfo {author} {\bibfnamefont {K.}~\bibnamefont
  {Landsteiner}}, \ and\ \bibinfo {author} {\bibfnamefont {L.}~\bibnamefont
  {Melgar}},\ }\href {\doibase 10.1103/PhysRevD.90.126004} {\bibfield
  {journal} {\bibinfo  {journal} {Phys. Rev. D}\ }\textbf {\bibinfo {volume}
  {90}},\ \bibinfo {pages} {126004} (\bibinfo {year} {2014})},\ \Eprint
  {http://arxiv.org/abs/1407.8162} {arXiv:1407.8162 [hep-th]} \BibitemShut
  {NoStop}%
\bibitem [{\citenamefont {Fukushima}\ \emph {et~al.}(2008)\citenamefont
  {Fukushima}, \citenamefont {Kharzeev},\ and\ \citenamefont
  {Warringa}}]{Fukushima:2008xe}%
  \BibitemOpen
  \bibfield  {author} {\bibinfo {author} {\bibfnamefont {K.}~\bibnamefont
  {Fukushima}}, \bibinfo {author} {\bibfnamefont {D.~E.}\ \bibnamefont
  {Kharzeev}}, \ and\ \bibinfo {author} {\bibfnamefont {H.~J.}\ \bibnamefont
  {Warringa}},\ }\href {\doibase 10.1103/PhysRevD.78.074033} {\bibfield
  {journal} {\bibinfo  {journal} {Phys. Rev. D}\ }\textbf {\bibinfo {volume}
  {78}},\ \bibinfo {pages} {074033} (\bibinfo {year} {2008})},\ \Eprint
  {http://arxiv.org/abs/0808.3382} {arXiv:0808.3382 [hep-ph]} \BibitemShut
  {NoStop}%
\bibitem [{\citenamefont {Gorbar}\ \emph {et~al.}(2014)\citenamefont {Gorbar},
  \citenamefont {Miransky}, \citenamefont {Shovkovy},\ and\ \citenamefont
  {Sukhachov}}]{Gorbar:2014qta}%
  \BibitemOpen
  \bibfield  {author} {\bibinfo {author} {\bibfnamefont {E.~V.}\ \bibnamefont
  {Gorbar}}, \bibinfo {author} {\bibfnamefont {V.~A.}\ \bibnamefont
  {Miransky}}, \bibinfo {author} {\bibfnamefont {I.~A.}\ \bibnamefont
  {Shovkovy}}, \ and\ \bibinfo {author} {\bibfnamefont {P.~O.}\ \bibnamefont
  {Sukhachov}},\ }\href {\doibase 10.1103/PhysRevB.90.115131} {\bibfield
  {journal} {\bibinfo  {journal} {Phys. Rev. B}\ }\textbf {\bibinfo {volume}
  {90}},\ \bibinfo {pages} {115131} (\bibinfo {year} {2014})},\ \Eprint
  {http://arxiv.org/abs/1407.1323} {arXiv:1407.1323 [cond-mat.str-el]}
  \BibitemShut {NoStop}%
\bibitem [{\citenamefont {Schlichting}\ and\ \citenamefont
  {Sharma}(2022)}]{Schlichting:2022fjc}%
  \BibitemOpen
  \bibfield  {author} {\bibinfo {author} {\bibfnamefont {S.}~\bibnamefont
  {Schlichting}}\ and\ \bibinfo {author} {\bibfnamefont {S.}~\bibnamefont
  {Sharma}},\ }\href@noop {} {\  (\bibinfo {year} {2022})},\ \Eprint
  {http://arxiv.org/abs/2211.11365} {arXiv:2211.11365 [hep-ph]} \BibitemShut
  {NoStop}%
\bibitem [{\citenamefont {Figueroa}\ \emph {et~al.}(2019)\citenamefont
  {Figueroa}, \citenamefont {Florio},\ and\ \citenamefont
  {Shaposhnikov}}]{Figueroa:2019jsi}%
  \BibitemOpen
  \bibfield  {author} {\bibinfo {author} {\bibfnamefont {D.~G.}\ \bibnamefont
  {Figueroa}}, \bibinfo {author} {\bibfnamefont {A.}~\bibnamefont {Florio}}, \
  and\ \bibinfo {author} {\bibfnamefont {M.}~\bibnamefont {Shaposhnikov}},\
  }\href {\doibase 10.1007/JHEP10(2019)142} {\bibfield  {journal} {\bibinfo
  {journal} {JHEP}\ }\textbf {\bibinfo {volume} {10}},\ \bibinfo {pages} {142}
  (\bibinfo {year} {2019})},\ \Eprint {http://arxiv.org/abs/1904.11892}
  {arXiv:1904.11892 [hep-th]} \BibitemShut {NoStop}%
\bibitem [{\citenamefont {Mace}\ \emph {et~al.}(2016)\citenamefont {Mace},
  \citenamefont {Schlichting},\ and\ \citenamefont
  {Venugopalan}}]{Mace:2016svc}%
  \BibitemOpen
  \bibfield  {author} {\bibinfo {author} {\bibfnamefont {M.}~\bibnamefont
  {Mace}}, \bibinfo {author} {\bibfnamefont {S.}~\bibnamefont {Schlichting}}, \
  and\ \bibinfo {author} {\bibfnamefont {R.}~\bibnamefont {Venugopalan}},\
  }\href {\doibase 10.1103/PhysRevD.93.074036} {\bibfield  {journal} {\bibinfo
  {journal} {Phys. Rev. D}\ }\textbf {\bibinfo {volume} {93}},\ \bibinfo
  {pages} {074036} (\bibinfo {year} {2016})},\ \Eprint
  {http://arxiv.org/abs/1601.07342} {arXiv:1601.07342 [hep-ph]} \BibitemShut
  {NoStop}%
\bibitem [{\citenamefont {Akamatsu}\ and\ \citenamefont
  {Yamamoto}(2013)}]{Akamatsu:2013pjd}%
  \BibitemOpen
  \bibfield  {author} {\bibinfo {author} {\bibfnamefont {Y.}~\bibnamefont
  {Akamatsu}}\ and\ \bibinfo {author} {\bibfnamefont {N.}~\bibnamefont
  {Yamamoto}},\ }\href {\doibase 10.1103/PhysRevLett.111.052002} {\bibfield
  {journal} {\bibinfo  {journal} {Phys. Rev. Lett.}\ }\textbf {\bibinfo
  {volume} {111}},\ \bibinfo {pages} {052002} (\bibinfo {year} {2013})},\
  \Eprint {http://arxiv.org/abs/1302.2125} {arXiv:1302.2125 [nucl-th]}
  \BibitemShut {NoStop}%
\bibitem [{\citenamefont {Hirono}\ \emph {et~al.}(2015)\citenamefont {Hirono},
  \citenamefont {Kharzeev},\ and\ \citenamefont {Yin}}]{Hirono:2015rla}%
  \BibitemOpen
  \bibfield  {author} {\bibinfo {author} {\bibfnamefont {Y.}~\bibnamefont
  {Hirono}}, \bibinfo {author} {\bibfnamefont {D.}~\bibnamefont {Kharzeev}}, \
  and\ \bibinfo {author} {\bibfnamefont {Y.}~\bibnamefont {Yin}},\ }\href
  {\doibase 10.1103/PhysRevD.92.125031} {\bibfield  {journal} {\bibinfo
  {journal} {Phys. Rev. D}\ }\textbf {\bibinfo {volume} {92}},\ \bibinfo
  {pages} {125031} (\bibinfo {year} {2015})},\ \Eprint
  {http://arxiv.org/abs/1509.07790} {arXiv:1509.07790 [hep-th]} \BibitemShut
  {NoStop}%
\bibitem [{\citenamefont {McLerran}\ \emph {et~al.}(1991)\citenamefont
  {McLerran}, \citenamefont {Mottola},\ and\ \citenamefont
  {Shaposhnikov}}]{McLerran:1990de}%
  \BibitemOpen
  \bibfield  {author} {\bibinfo {author} {\bibfnamefont {L.~D.}\ \bibnamefont
  {McLerran}}, \bibinfo {author} {\bibfnamefont {E.}~\bibnamefont {Mottola}}, \
  and\ \bibinfo {author} {\bibfnamefont {M.~E.}\ \bibnamefont {Shaposhnikov}},\
  }\href {\doibase 10.1103/PhysRevD.43.2027} {\bibfield  {journal} {\bibinfo
  {journal} {Phys. Rev. D}\ }\textbf {\bibinfo {volume} {43}},\ \bibinfo
  {pages} {2027} (\bibinfo {year} {1991})}\BibitemShut {NoStop}%
\bibitem [{\citenamefont {Moore}\ and\ \citenamefont
  {Tassler}(2011)}]{Moore:2010jd}%
  \BibitemOpen
  \bibfield  {author} {\bibinfo {author} {\bibfnamefont {G.~D.}\ \bibnamefont
  {Moore}}\ and\ \bibinfo {author} {\bibfnamefont {M.}~\bibnamefont
  {Tassler}},\ }\href {\doibase 10.1007/JHEP02(2011)105} {\bibfield  {journal}
  {\bibinfo  {journal} {JHEP}\ }\textbf {\bibinfo {volume} {02}},\ \bibinfo
  {pages} {105} (\bibinfo {year} {2011})},\ \Eprint
  {http://arxiv.org/abs/1011.1167} {arXiv:1011.1167 [hep-ph]} \BibitemShut
  {NoStop}%
\bibitem [{\citenamefont {Arnold}\ and\ \citenamefont
  {McLerran}(1988)}]{Arnold:1987zg}%
  \BibitemOpen
  \bibfield  {author} {\bibinfo {author} {\bibfnamefont {P.~B.}\ \bibnamefont
  {Arnold}}\ and\ \bibinfo {author} {\bibfnamefont {L.~D.}\ \bibnamefont
  {McLerran}},\ }\href {\doibase 10.1103/PhysRevD.37.1020} {\bibfield
  {journal} {\bibinfo  {journal} {Phys. Rev. D}\ }\textbf {\bibinfo {volume}
  {37}},\ \bibinfo {pages} {1020} (\bibinfo {year} {1988})}\BibitemShut
  {NoStop}%
\bibitem [{\citenamefont {Basar}\ and\ \citenamefont
  {Kharzeev}(2012)}]{Basar:2012gh}%
  \BibitemOpen
  \bibfield  {author} {\bibinfo {author} {\bibfnamefont {G.}~\bibnamefont
  {Basar}}\ and\ \bibinfo {author} {\bibfnamefont {D.~E.}\ \bibnamefont
  {Kharzeev}},\ }\href {\doibase 10.1103/PhysRevD.85.086012} {\bibfield
  {journal} {\bibinfo  {journal} {Phys. Rev. D}\ }\textbf {\bibinfo {volume}
  {85}},\ \bibinfo {pages} {086012} (\bibinfo {year} {2012})},\ \Eprint
  {http://arxiv.org/abs/1202.2161} {arXiv:1202.2161 [hep-th]} \BibitemShut
  {NoStop}%
\bibitem [{\citenamefont {Altenkort}\ \emph {et~al.}(2021)\citenamefont
  {Altenkort}, \citenamefont {Eller}, \citenamefont {Kaczmarek}, \citenamefont
  {Mazur}, \citenamefont {Moore},\ and\ \citenamefont
  {Shu}}]{Altenkort:2020axj}%
  \BibitemOpen
  \bibfield  {author} {\bibinfo {author} {\bibfnamefont {L.}~\bibnamefont
  {Altenkort}}, \bibinfo {author} {\bibfnamefont {A.~M.}\ \bibnamefont
  {Eller}}, \bibinfo {author} {\bibfnamefont {O.}~\bibnamefont {Kaczmarek}},
  \bibinfo {author} {\bibfnamefont {L.}~\bibnamefont {Mazur}}, \bibinfo
  {author} {\bibfnamefont {G.~D.}\ \bibnamefont {Moore}}, \ and\ \bibinfo
  {author} {\bibfnamefont {H.-T.}\ \bibnamefont {Shu}},\ }\href {\doibase
  10.1103/PhysRevD.103.114513} {\bibfield  {journal} {\bibinfo  {journal}
  {Phys. Rev. D}\ }\textbf {\bibinfo {volume} {103}},\ \bibinfo {pages}
  {114513} (\bibinfo {year} {2021})},\ \Eprint
  {http://arxiv.org/abs/2012.08279} {arXiv:2012.08279 [hep-lat]} \BibitemShut
  {NoStop}%
\bibitem [{\citenamefont {Son}\ and\ \citenamefont
  {Surowka}(2009)}]{Son:2009tf}%
  \BibitemOpen
  \bibfield  {author} {\bibinfo {author} {\bibfnamefont {D.~T.}\ \bibnamefont
  {Son}}\ and\ \bibinfo {author} {\bibfnamefont {P.}~\bibnamefont {Surowka}},\
  }\href {\doibase 10.1103/PhysRevLett.103.191601} {\bibfield  {journal}
  {\bibinfo  {journal} {Phys. Rev. Lett.}\ }\textbf {\bibinfo {volume} {103}},\
  \bibinfo {pages} {191601} (\bibinfo {year} {2009})},\ \Eprint
  {http://arxiv.org/abs/0906.5044} {arXiv:0906.5044 [hep-th]} \BibitemShut
  {NoStop}%
\bibitem [{\citenamefont {Arnold}\ \emph {et~al.}(2000)\citenamefont {Arnold},
  \citenamefont {Moore},\ and\ \citenamefont {Yaffe}}]{Arnold:2000dr}%
  \BibitemOpen
  \bibfield  {author} {\bibinfo {author} {\bibfnamefont {P.~B.}\ \bibnamefont
  {Arnold}}, \bibinfo {author} {\bibfnamefont {G.~D.}\ \bibnamefont {Moore}}, \
  and\ \bibinfo {author} {\bibfnamefont {L.~G.}\ \bibnamefont {Yaffe}},\ }\href
  {\doibase 10.1088/1126-6708/2000/11/001} {\bibfield  {journal} {\bibinfo
  {journal} {JHEP}\ }\textbf {\bibinfo {volume} {11}},\ \bibinfo {pages} {001}
  (\bibinfo {year} {2000})},\ \Eprint {http://arxiv.org/abs/hep-ph/0010177}
  {arXiv:hep-ph/0010177} \BibitemShut {NoStop}%
\bibitem [{\citenamefont {Bernhard}\ \emph {et~al.}(2019)\citenamefont
  {Bernhard}, \citenamefont {Moreland},\ and\ \citenamefont
  {Bass}}]{Bernhard:2019bmu}%
  \BibitemOpen
  \bibfield  {author} {\bibinfo {author} {\bibfnamefont {J.~E.}\ \bibnamefont
  {Bernhard}}, \bibinfo {author} {\bibfnamefont {J.~S.}\ \bibnamefont
  {Moreland}}, \ and\ \bibinfo {author} {\bibfnamefont {S.~A.}\ \bibnamefont
  {Bass}},\ }\href {\doibase 10.1038/s41567-019-0611-8} {\bibfield  {journal}
  {\bibinfo  {journal} {Nature Phys.}\ }\textbf {\bibinfo {volume} {15}},\
  \bibinfo {pages} {1113} (\bibinfo {year} {2019})}\BibitemShut {NoStop}%
\bibitem [{\citenamefont {Sadofyev}\ and\ \citenamefont
  {Isachenkov}(2011)}]{Sadofyev:2010pr}%
  \BibitemOpen
  \bibfield  {author} {\bibinfo {author} {\bibfnamefont {A.~V.}\ \bibnamefont
  {Sadofyev}}\ and\ \bibinfo {author} {\bibfnamefont {M.~V.}\ \bibnamefont
  {Isachenkov}},\ }\href {\doibase 10.1016/j.physletb.2011.02.041} {\bibfield
  {journal} {\bibinfo  {journal} {Phys. Lett. B}\ }\textbf {\bibinfo {volume}
  {697}},\ \bibinfo {pages} {404} (\bibinfo {year} {2011})},\ \Eprint
  {http://arxiv.org/abs/1010.1550} {arXiv:1010.1550 [hep-th]} \BibitemShut
  {NoStop}%
\bibitem [{\citenamefont {Huang}\ and\ \citenamefont
  {Liao}(2013)}]{Huang:2013iia}%
  \BibitemOpen
  \bibfield  {author} {\bibinfo {author} {\bibfnamefont {X.-G.}\ \bibnamefont
  {Huang}}\ and\ \bibinfo {author} {\bibfnamefont {J.}~\bibnamefont {Liao}},\
  }\href {\doibase 10.1103/PhysRevLett.110.232302} {\bibfield  {journal}
  {\bibinfo  {journal} {Phys. Rev. Lett.}\ }\textbf {\bibinfo {volume} {110}},\
  \bibinfo {pages} {232302} (\bibinfo {year} {2013})},\ \Eprint
  {http://arxiv.org/abs/1303.7192} {arXiv:1303.7192 [nucl-th]} \BibitemShut
  {NoStop}%
\bibitem [{\citenamefont {Onsager}(1931)}]{Onsager:1931jfa}%
  \BibitemOpen
  \bibfield  {author} {\bibinfo {author} {\bibfnamefont {L.}~\bibnamefont
  {Onsager}},\ }\href {\doibase 10.1103/physrev.37.405} {\bibfield  {journal}
  {\bibinfo  {journal} {Phys. Rev.}\ }\textbf {\bibinfo {volume} {37}},\
  \bibinfo {pages} {405} (\bibinfo {year} {1931})}\BibitemShut {NoStop}%
\bibitem [{\citenamefont {Son}\ and\ \citenamefont
  {Zhitnitsky}(2004)}]{Son:2004tq}%
  \BibitemOpen
  \bibfield  {author} {\bibinfo {author} {\bibfnamefont {D.~T.}\ \bibnamefont
  {Son}}\ and\ \bibinfo {author} {\bibfnamefont {A.~R.}\ \bibnamefont
  {Zhitnitsky}},\ }\href {\doibase 10.1103/PhysRevD.70.074018} {\bibfield
  {journal} {\bibinfo  {journal} {Phys. Rev. D}\ }\textbf {\bibinfo {volume}
  {70}},\ \bibinfo {pages} {074018} (\bibinfo {year} {2004})},\ \Eprint
  {http://arxiv.org/abs/hep-ph/0405216} {arXiv:hep-ph/0405216} \BibitemShut
  {NoStop}%
\bibitem [{\citenamefont {Metlitski}\ and\ \citenamefont
  {Zhitnitsky}(2005)}]{Metlitski:2005pr}%
  \BibitemOpen
  \bibfield  {author} {\bibinfo {author} {\bibfnamefont {M.~A.}\ \bibnamefont
  {Metlitski}}\ and\ \bibinfo {author} {\bibfnamefont {A.~R.}\ \bibnamefont
  {Zhitnitsky}},\ }\href {\doibase 10.1103/PhysRevD.72.045011} {\bibfield
  {journal} {\bibinfo  {journal} {Phys. Rev. D}\ }\textbf {\bibinfo {volume}
  {72}},\ \bibinfo {pages} {045011} (\bibinfo {year} {2005})},\ \Eprint
  {http://arxiv.org/abs/hep-ph/0505072} {arXiv:hep-ph/0505072} \BibitemShut
  {NoStop}%
\bibitem [{\citenamefont {Kharzeev}\ and\ \citenamefont
  {Yee}(2011)}]{Kharzeev:2010gd}%
  \BibitemOpen
  \bibfield  {author} {\bibinfo {author} {\bibfnamefont {D.~E.}\ \bibnamefont
  {Kharzeev}}\ and\ \bibinfo {author} {\bibfnamefont {H.-U.}\ \bibnamefont
  {Yee}},\ }\href {\doibase 10.1103/PhysRevD.83.085007} {\bibfield  {journal}
  {\bibinfo  {journal} {Phys. Rev. D}\ }\textbf {\bibinfo {volume} {83}},\
  \bibinfo {pages} {085007} (\bibinfo {year} {2011})},\ \Eprint
  {http://arxiv.org/abs/1012.6026} {arXiv:1012.6026 [hep-th]} \BibitemShut
  {NoStop}%
\bibitem [{\citenamefont {Skokov}\ \emph {et~al.}(2009)\citenamefont {Skokov},
  \citenamefont {Illarionov},\ and\ \citenamefont {Toneev}}]{Skokov:2009qp}%
  \BibitemOpen
  \bibfield  {author} {\bibinfo {author} {\bibfnamefont {V.}~\bibnamefont
  {Skokov}}, \bibinfo {author} {\bibfnamefont {A.~Y.}\ \bibnamefont
  {Illarionov}}, \ and\ \bibinfo {author} {\bibfnamefont {V.}~\bibnamefont
  {Toneev}},\ }\href {\doibase 10.1142/S0217751X09047570} {\bibfield  {journal}
  {\bibinfo  {journal} {Int. J. Mod. Phys. A}\ }\textbf {\bibinfo {volume}
  {24}},\ \bibinfo {pages} {5925} (\bibinfo {year} {2009})},\ \Eprint
  {http://arxiv.org/abs/0907.1396} {arXiv:0907.1396 [nucl-th]} \BibitemShut
  {NoStop}%
\bibitem [{\citenamefont {Koch}\ \emph
  {et~al.}(2017{\natexlab{b}})\citenamefont {Koch}, \citenamefont
  {Schlichting}, \citenamefont {Skokov}, \citenamefont {Sorensen},
  \citenamefont {Thomas}, \citenamefont {Voloshin}, \citenamefont {Wang},\ and\
  \citenamefont {Yee}}]{Koch:2016pzl}%
  \BibitemOpen
  \bibfield  {author} {\bibinfo {author} {\bibfnamefont {V.}~\bibnamefont
  {Koch}}, \bibinfo {author} {\bibfnamefont {S.}~\bibnamefont {Schlichting}},
  \bibinfo {author} {\bibfnamefont {V.}~\bibnamefont {Skokov}}, \bibinfo
  {author} {\bibfnamefont {P.}~\bibnamefont {Sorensen}}, \bibinfo {author}
  {\bibfnamefont {J.}~\bibnamefont {Thomas}}, \bibinfo {author} {\bibfnamefont
  {S.}~\bibnamefont {Voloshin}}, \bibinfo {author} {\bibfnamefont
  {G.}~\bibnamefont {Wang}}, \ and\ \bibinfo {author} {\bibfnamefont {H.-U.}\
  \bibnamefont {Yee}},\ }\href {\doibase 10.1088/1674-1137/41/7/072001}
  {\bibfield  {journal} {\bibinfo  {journal} {Chin. Phys. C}\ }\textbf
  {\bibinfo {volume} {41}},\ \bibinfo {pages} {072001} (\bibinfo {year}
  {2017}{\natexlab{b}})},\ \Eprint {http://arxiv.org/abs/1608.00982}
  {arXiv:1608.00982 [nucl-th]} \BibitemShut {NoStop}%
\end{thebibliography}%

\appendix

\section{Divergence of entropy current, magnetic field, vorticity}\label{app:divergences}

Let us first derive the divergence of the entropy current. The viscous correction to the entropy current in first order hydrodynamics is 
\begin{align}
S^{\mu} = \frac{(\epsilon+P)}{T} u^{\mu} - \frac{\mu_{V,f}}{T} j_{V,f}^{\mu} - \frac{\mu_{A,f}}{T} j_{A,f}^{\mu}.
\label{eq:entcurrent}
\end{align}
Immediately, the divergence takes the form
\begin{align}
\partial_{\mu}S^{\mu}=&\partial_{\mu}\left(\frac{(\epsilon+P)}{T} u^{\mu} \right) 
-\partial_{\mu}\bigg(\frac{\mu_{V,f}}{T}n_{V,f} u^{\mu}\nonumber\\
&+\frac{\mu_{A,f}}{T}n_{A,f} u^{\mu}+\frac{\mu_{V,f}}{T}\nu_{V,f}^{\mu} +\frac{\mu_{A,f}}{T}\nu_{A,f}^{\mu}\bigg).\quad
\label{eq:sdiv1}
\end{align}
We focus on the first term on the right hand side of Eq. \eqref{eq:sdiv1}. From the longitudinal projection on the divergence of the energy-momentum tensor, we have
\begin{align}
\partial_{\mu} (\epsilon+P) u^{\mu} = u^{\mu}\partial_{\mu} P - \tau^{\mu\nu} \left(\partial_{\mu} u_{\nu}\right)  - u_{\nu} \left( \partial_{\mu} T^{\mu\nu} \right).\quad\ \label{eq:longproj}
\end{align}
Now, using the thermodynamic relations $Ts=(\epsilon+P)-\mu_{A,f} n_{A,f}-\mu_{V,f}n_{V,f}$ and $dP=sdT+n _{A,f} d\mu_{A,f}+n _{V,f} d\mu_{V,f}$, we can express
\begin{align}
\frac{1}{T} u^{\mu}\partial_{\mu} P =& -(\epsilon+P) u^{\mu} \left(\partial_{\mu} \frac{1}{T} \right) + n_{V,f} u^{\mu} \left(\partial_{\mu} \frac{\mu_{V,f}}{T} \right) \nonumber\\
&+ n_{A,f} u^{\mu} \left(\partial_{\mu} \frac{\mu_{A,f}}{T} \right).\quad \label{eq:thankunext}
\end{align}
Dividing Eq. (\ref{eq:longproj}) by $T$, and then combining with Eq. (\ref{eq:thankunext}), we obtain 
\begin{align}
\partial_{\mu} \left(\frac{\epsilon+P}{T} u^{\mu}   \right) =& -  \frac{1}{T} \tau^{\mu\nu} \left(\partial_{\mu} u_{\nu}\right) - \frac{1}{T} u_{\nu} \left(\partial_{\mu} T^{\mu\nu} \right)\nonumber\\
&+ \sum_{j=V,A} n_{j,f} u^{\mu} \left(\partial_{\mu} \frac{\mu_{j,f}}{T} \right).
\end{align}
We substitute this result into Eq. (\ref{eq:sdiv1}), and obtain 
\begin{align}
\partial_{\mu} S^{\mu} =&  - \frac{1}{T} \tau^{\mu\nu} \left(\partial_{\mu} u_{\nu}\right)  -\frac{1}{T} u_{\nu} \left(\partial_{\mu} T^{\mu\nu}\right)  \nonumber\\
&- (j^{\mu}_{V,f}- n_{V,f} u^{\mu}) \left(\partial_{\mu} \frac{\mu_{V,f}}{T} \right) - \frac{\mu_{V,f}}{T} \left( \partial_{\mu} J_{V,f}^{\mu} \right). \nonumber \\
&- (j^{\mu}_{A,f}- n_{A,f} u^{\mu}) \left(\partial_{\mu} \frac{\mu_{A,f}}{T} \right) - \frac{\mu_{A,f}}{T} \left( \partial_{\mu} J_{A,f}^{\mu} \right).\nonumber\\
\label{eq:sdiv2}
\end{align}
Identifying  $j^{\mu}_{i}- n_{i} u^{\mu}= \nu^{\mu}_{i}$ , we evaluate 
\begin{align}
u_{\nu} \left(\partial_{\mu} T^{\mu\nu}\right)= - \sum_{f} e q_f E_{\lambda} \left(j^{\lambda}_{V,f} - n_{V,f} u^{\lambda}\right),\
\end{align}
where in the last step we used the fact that $E^{\mu}$ is a spacelike vector (i.e $u_{\mu}E^{\mu}=0$). After making the appropriate substitutions into Eq. (\ref{eq:sdiv2}), it is straightforward to find
\begin{align}
\partial_{\mu} S^{\mu} =&  - \frac{1}{T} \tau^{\mu\nu} \left(\partial_{\mu} u_{\nu}\right) - \nu^{\mu}_{V,f}\left( \left(\partial_{\mu} \frac{\mu_{V,f}}{T} \right) - \frac{eq_f }{T} E_{\mu} \right)\nonumber\\ 
& - \nu^{\mu}_{A,f} \left(\partial_{\mu} \frac{\mu_{A,f}}{T} \right)- \frac{\mu_{A,f}}{T} \left( \partial_{\mu} J_{A,f}^{\mu} \right).
\end{align}
Then, we note that the most general modification we can make to the entropy current Eq. (\ref{eq:entcurrent}) in the presence of an external magnetic and vorticity field is
\begin{align}
S_{\rm ext}^{\mu}=& D_BB^{\mu}+D_{\omega}\omega^{\mu},
\end{align}
with divergence
\begin{align}
\partial_{\mu}S_{\rm ext}^{\mu}=& \left(\partial_{\mu}D_B\right)B^{\mu}+D_B\partial_{\mu}B^{\mu}+\left(\partial_{\mu}D_{\omega}\right)\omega^{\mu}\nonumber \\
&+D_{\omega}\partial_{\mu}\omega^{\mu}.
\label{eq:mods}
\end{align}
Hence, the total divergence of the entropy current is expressed in Eq. (\ref{eq:divs}).

Now, we can derive identities for the divergences of vorticity $\omega^{\mu}$ and magnetic field $B^{\mu}$ found in Eq. (\ref{eq:mods}) using the ideal hydrodynamic equations. We first note that both fields $B^{\mu}$ and $\omega^{\mu}$ can be expressed in an analogous way in terms of the components of the tensors $F_{\alpha\beta}=\partial_{\alpha} A_{\beta} - \partial_{\beta} A_{\alpha}$ and $\Omega_{\alpha\beta}=\frac{1}{2}\epsilon^{\mu\nu\alpha\beta} \partial_{\alpha} u_{\beta}$ according to
\begin{align}
F_{\alpha\beta} &= u_{\alpha} E_{\beta} - u_{\beta} E_{\alpha} - \epsilon_{\alpha\beta\gamma\delta} u^{\gamma} B^{\delta}\;, \\
\Omega_{\alpha\beta} &= u_{\lambda} \omega_{\beta} - u_{\beta} \omega_{\alpha} -\epsilon_{\alpha\beta\gamma\delta} u^{\gamma} a^{\delta}\;,
\end{align}
where 
\begin{align}
a^{\mu}=\frac{1}{2} \epsilon^{\mu\nu\alpha\beta} u_{\nu} \Omega_{\alpha\beta} = \frac{1}{2} u_{\nu} \partial^{\nu} u^{\mu}
\end{align}
is the acceleration of the fluid. Since for sufficiently smooth fields $\partial_{\mu} \Omega^{\mu\nu}=0$, we can then express the derivatives as
\begin{align}
\partial_{\mu} B^{\mu} &= \frac{1}{2} \epsilon^{\mu\nu\alpha\beta} (\partial_{\mu} u_{\nu} ) F_{\alpha\beta}= - 2\omega_{\mu}E^{\mu} + 2 a_{\mu} B^{\mu}, \qquad\\
\partial_{\mu} \omega^{\mu} &= \partial_{\mu} \left(\Omega^{\mu\nu} u_{\nu} \right) = (\partial_{\mu} u_{\nu} ) \Omega^{\mu\nu}= 4 a_{\mu} \omega^{\mu}\;.
\end{align}
By transversely projecting the energy-momentum conservation equation $\Delta^{\alpha}_{~\nu} \partial_{\mu} T^{\mu\nu}$, one obtains
\begin{align}
(\epsilon+p) u^{\mu}\partial_{\mu} u^{\alpha} =&- \Delta^{\alpha}_{~\nu} \partial^{\nu} P + \sum_{f} eq_{f} \Delta^{\alpha}_{~\nu} F^{\nu\lambda} j_{\lambda}^{V,f} \nonumber\\
&+\Delta^{\alpha}_{~\nu}\partial_{\mu} \tau^{\mu\nu} ,
\end{align}
which upon keeping only terms linear in gradients 
becomes
\begin{align}
(\epsilon+P) u^{\mu}\partial_{\mu} u^{\alpha} =& - \Delta^{\alpha}_{~\nu} \partial^{\nu} P + \sum_{f} e q_{f} n^{V,f}  E^{\alpha}  \nonumber \\
&+ \mathcal{O}(\partial^2).\qquad
\end{align}
Collecting everything, one obtains the identities
\begin{align}
\partial_{\mu} B^{\mu} =& -2\omega_{\mu}E^{\mu} -\frac{B^{\mu}}{\epsilon+P} \left(\left(\partial_{\mu}P\right)  -\sum_fe q_{f} n_{V,f}  E_{\mu}  \right), \nonumber\\
&\\
\partial_{\mu} \omega^{\mu} =& -\frac{2 \omega^{\mu}}{e+p} \left(\left(\partial_{\mu}P\right)  -\sum_fe q_{f} n_{V,f}  E_{\mu}  \right)\;.
\end{align}

\section{Constraints on transport coefficients}\label{app:transport}

In order to determine the constraints on the chiral coefficients, following \cite{Son:2009tf}, we use the identities derived in Appendix \ref{app:divergences} that follow from the ideal hydrodynamic equations:
\begin{align}
\partial_{\mu}\omega^{\mu} =& -\frac{2\omega^{\mu}}{\epsilon + P} \left( \partial_{\mu}P - \sum_{f}e q_f n_{V,f} E_{\mu}\right), \label{eq:dmu_vort}\\
\partial_{\mu} B^{\mu} =& -2\omega^{\mu}E_{\mu} + \frac{1}{\epsilon + P}\bigg(-B^{\mu}\partial_{\mu} P\nonumber\\
&+\sum_{f}e q_f  n_{V,f}  E^{\mu}B_{\mu} \bigg). \label{eq:dmu_b}
\end{align}
By inserting Eqs.~(\ref{eq:dmu_vort},\ref{eq:dmu_b}) and the expanded forms of $\nu_{V/A}^{\mu}$ (Eqs. \ref{eq:nuv}, \ref{eq:nua}) into Eq.~(\ref{eq:divs}), one then finds various contributions to the divergence of the entropy current that are proportional to either $\omega^{\mu}$, $B^{\mu}$, $E_{\mu}\omega^{\mu}$, or $E_{\mu}B^{\mu}$. Since neither of these terms has a definite sign, in order to comply with a locally positive semi-definite entropy production, the combinations of coefficients multiplying them must vanish identically such that the effects associated with the coupling to $\omega^{\mu}$, $B^{\mu}$, $E_{\mu}\omega^{\mu}$, or $E_{\mu}B^{\mu}$ are in fact non-dissipative, yielding the equations
\begin{widetext}
\begin{align}
\left(\partial_{\mu} D_{\omega} -\xi_{A,f} \partial_{\mu}\frac{\mu_{A,f}}{T}-\xi_{V,f} \partial_{\mu}\frac{\mu_{V,f}}{T} - \frac{2D_{\omega}}{\epsilon + P} \partial_{\mu}P \right)  \left( \omega^{\mu} \right)&= 0,\qquad \label{eq:1}\\
\left(\frac{e q_f \xi_{V,f}}{T} - 2D_B + \frac{2D_{\omega}}{\epsilon +P}\sum_fe q_f n_{V,f}\right)  \left( E_{\mu} \omega^{\mu} \right)&= 0, \label{eq:2} \\
\left(\partial_{\mu}D_B- e q_f \sigma_{AB}^{f}\partial_{\mu}\frac{\mu_{A,f}}{T}- e q_f \sigma_{VB}^{f}\partial_{\mu}\frac{\mu_{V,f}}{T} - \frac{D_B}{\epsilon + P} \partial_{\mu}P\right) \left( B^{\mu} \right)&= 0, \label{eq:3}\\
\left(\sum_{f} \frac{e^2 q_f^2 \sigma_{VB}^{f}}{T} -C \sum_{f} \frac{\mu_{A,f}}{T} (eq_f)^2 + \frac{D_B}{\epsilon + P} \sum_fe q_f  n_{V,f} \right) \left( E_{\mu} B^{\mu} \right)&= 0. \label{eq:4}
\end{align}
\end{widetext}
In order to evaluate these constraints more explicitly, it is convenient to then switch variables from $T$ and $\mu_{A/V,f}$ to $\overline{\mu}_{A/V,f} \equiv \mu_{A/V,f}/T$ and $P$. Based on the thermodynamic relations $Ts = \epsilon + P - \mu_{V,f} n_{V,f}- \mu_{A,f} n_{A,f}$ and $dP = sdT + n_{V,f}d\mu_{V,f}+n_{A,f}d\mu_{A,f}$, one finds the relations 
\begin{align}
\left(\frac{\partial T}{\partial P}\right)_{\overline{\mu}_{i}} = \frac{T}{\epsilon + P}, \label{eq:TDrels}\; \quad
\left(\frac{\partial T}{\partial \overline{\mu}_{i}}\right)_{P,\overline{\mu}_j} = -\frac{n_{i}T^2}{\epsilon +P}. 
\end{align}
Expressing the derivatives of the various coefficients in Eq. (\ref{eq:1}) and Eq. (\ref{eq:3}) as 
\begin{align} 
\partial_{\mu} X =& \left(\frac{\partial X}{\partial P}\right) \partial_{\mu}P + \left(\frac{\partial X}{\partial \overline{\mu}_{V,f}}\right) \partial_{\mu} \overline{\mu}_{V,f} \nonumber\\
& + \left(\frac{\partial X}{\partial \overline{\mu}_{A,f}}\right) \partial_{\mu} \overline{\mu}_{A,f},
\end{align} 
and exploiting the fact that variations of $P$ and $\bar{\mu}_{V/A,f}$ are independent of each other, one finds that Eq. (\ref{eq:1}) splits into $2N_{f}+1$ equations:
\begin{align} 
\frac{\partial D_{\omega}}{\partial \overline{\mu}_{V,f}} = \xi_V^{f}\;, \quad
\frac{\partial D_{\omega}}{\partial \overline{\mu}_{A,f}} = \xi_A^{f}\;, \quad
\frac{\partial D_{\omega}}{\partial P} = \frac{2 D_{\omega}}{\epsilon + P}.
\end{align}
Based on Eq. (\ref{eq:TDrels}), one then concludes that the solutions for Eq. (\ref{eq:1}) are of the form
\begin{align}
D_{\omega} &= T^2 f_{\omega}(\overline{\mu}_{V,f},\overline{\mu}_{A,f}),\\
\xi_V^{f} &= \frac{\partial}{\partial \overline{\mu}_{V,f}}\left( T^2  f_{\omega}(\overline{\mu}_{V,f},\overline{\mu}_{A,f})\right),\\
\xi_A^{f} &= \frac{\partial}{\partial \overline{\mu}_{A,f}}\left( T^2  f_{\omega}(\overline{\mu}_{V,f},\overline{\mu}_{A,f})\right),
\end{align}
in which $f_{\omega}(\overline{\mu}_{V,f},\overline{\mu}_{A,f})$ is a hitherto arbitrary function of $\overline{\mu}_{V,f}$ and $\overline{\mu}_{A,f}$ . Similarly, Eq. (\ref{eq:3}) also splits into $2N_{f}+1$ equations,
\begin{align}
\frac{\partial D_{B}}{\partial \overline{\mu}_{V,f}} = e q_f \sigma_{VB}^{f}\;,\
\frac{\partial D_{B}}{\partial \overline{\mu}_{A,f}} = e q_f \sigma_{AB}^{f}\;, \ 
\frac{\partial D_{B}}{\partial P} =  \frac{D_{B}}{\epsilon + P},\nonumber \\
\end{align}
which with the help of Eq.~(\ref{eq:TDrels}) yields
\begin{align}
D_{B} &= T f_{B}(\overline{\mu}_{V,f},\overline{\mu}_{A,f})\;, \\
e q_f \sigma_{VB}^{f} &= \frac{\partial}{\partial \bar{\mu}_{V,f}}\left( T f_{B}(\overline{\mu}_{V,f},\overline{\mu}_{A,f})\right)\;, \\
e q_f \sigma_{AB}^{f} &= \frac{\partial}{\partial \bar{\mu}_{A,f}}\left( T  f_{B}(\overline{\mu}_{V,f},\overline{\mu}_{A,f})\right)\;.
\end{align}
By taking into account Eq. (\ref{eq:2}) and Eq. (\ref{eq:4}), one then finds
\begin{align}
\sum_{f} e q_f \frac{1}{2}\frac{\partial f_{\omega}}{\partial \overline{\mu}_{V,f}} = f_B(\overline{\mu}_{V,f},\overline{\mu}_{A,f})\;, \\
\sum_{f} e q_f \frac{\partial f_B}{\partial \overline{\mu}_{V,f}} = C \sum_{f} e^2q_f^2  \overline{\mu}_{A,f},\qquad\;.
\end{align}
Specifically, for the case of a single flavor ($N_{f}=1$), the functions $f_\omega$ and $f_B$ can then be obtained directly via integration
\begin{align}
f_B(\overline{\mu}_V,\overline{\mu}_A) &= e q_f C \overline{\mu}_A\overline{\mu}_V + g(\overline{\mu}_A) , \\  
f_{\omega}(\overline{\mu}_V,\overline{\mu}_A) &= C \overline{\mu}_V^2\overline{\mu}_A + \frac{\overline{\mu}_V}{e q_f}g(\overline{\mu}_A) + G(\overline{\mu}_A),
\end{align}
where $g(\overline{\mu}_A)$ and $G(\overline{\mu}_A)$ are hitherto arbitrary functions of $\overline{\mu}_A$.

Generalizing the single-flavor result to multiple independent flavors and dropping the unspecified contributions then yields
\begin{align}
f_B(\overline{\mu}_V,\overline{\mu}_A) &= C \sum_{f} e q_f  \overline{\mu}_{A,f}\overline{\mu}_{V,f}, \\  
f_{\omega}(\overline{\mu}_V,\overline{\mu}_A) &= C \sum_{f} \overline{\mu}_{V,f}^2\overline{\mu}_{A,f}.  
\end{align}


\section{Degenerate perturbation theory calculations for multi-flavor dynamics}\label{app:multiflavor}
\begin{widetext}
Below we explain the calculation of the eigenmodes in the two quark-flavor case. We focus for simplicity on the case $\chi_V=\chi_A=\chi$, where the matrix is symmetric and the calculations can be carried out in a familiar fashion. One finds that to leading order in the small $k$ limit, the matrix $M_{ab}^{N_f=2}$ (Eq. \ref{eq:mf_matrix}) becomes
\begin{align}
M^{N_f=2}_{k=0}&=
\begin{pmatrix}
0 & 0 & 0 & 0 \\
0 & \gsph & 0 & \gsph \\
0 & 0 & 0 & 0 \\
0 & \gsph & 0 & \gsph
\end{pmatrix},
\label{eq:k0matrix}
\end{align}
with eigenvalues
\begin{align}
    \lambda_1=2\gsph, \qquad \lambda_{2}=\lambda_3=\lambda_4=0.
\end{align}
We use degenerate perturbation theory to disentangle the three degenerate eigenvalues and determine the perturbations up to first order in $k$. By perturbing the matrix \eqref{eq:k0matrix} with the first order contributions from \eqref{eq:mf_matrix}, one obtains the first-order matrix,
\begin{align}
    M^{N_f=2}\bigg|_{\mathcal{O}(k)} =
    \begin{pmatrix}
        0 & ieq_u C\chi^{-1} \kt\cdot\Bt& 0 & 0 \\
        ieq_u C\chi^{-1} \kt\cdot\Bt & \gsph & 0 & \gsph\\
        0 & 0 & 0 & ieq_d C\chi^{-1} \kt\cdot\Bt \\
        0 & \gsph & ieq_d C\chi^{-1} \kt\cdot\Bt & \gsph
    \end{pmatrix}.
    \label{eq:orderkmat}
\end{align}
We then can choose an orthonormal basis for the leading order eigenvectors that diagonalizes the degenerate subspace,
\begin{align}
\et_{1} &= \frac{1}{\sqrt{2}} \left\{ 0,1,0,1\right\}, \\
\et_{2}&= \frac{1}{\sqrt{q_d^2+q_u^2}}\left\{q_d,0,q_u,0\right\},\\
\et_{3}&= \frac{1}{\sqrt{2}}\left\{\frac{-q_u}{\sqrt{q_d^2+q_u^2}},\frac{1}{\sqrt{2}},\frac{q_d}{\sqrt{q_d^2+q_u^2}},-\frac{1}{\sqrt{2}} \right\}\\
\et_{4}&= \frac{1}{\sqrt{2}}\left\{\frac{-q_u}{\sqrt{q_d^2+q_u^2}},-\frac{1}{\sqrt{2}},\frac{q_d}{\sqrt{q_d^2+q_u^2}},\frac{1}{\sqrt{2}} \right\}.
\label{eq:newbasis}
\end{align}
By projecting the matrix \eqref{eq:orderkmat} onto the leading order eigenvectors in Eq.~\eqref{eq:newbasis}, we obtain a matrix of the form
\begin{align}
    M^{N_f=2}\bigg|_{\mathcal{O}(k)} =
    \begin{pmatrix}
 2\gamma & k \vec{v}^{T} \\
 k \vec{v} & k D
    \end{pmatrix}
    \label{eq:orderkmat}
\end{align}
where the matrix $D$ describes the mixing between the degenerate leading order eigenvectors $(i,j=2,3,4)$,
\begin{align}
kD_{ij}=\et_i^{T}\left(M^{N_f=2}\bigg|_{\mathcal{O}(k)}-M^{N_f=2}_{k=0} \right)\et_j\;, \qquad  kD=
\begin{pmatrix} 
0&0&0 \\
0&-\frac{ieC\kt\cdot\Bt}{\chi\sqrt{2}}\sqrt{q_d^2+q_u^2}&0 \\
0&0& \frac{ieC\kt\cdot\Bt}{\chi\sqrt{2}}\sqrt{q_d^2+q_u^2}
\end{pmatrix},
\label{eq:diagpmat}
\end{align}
and the vector $\vec{v}$ describes the coupling between the degenerate eigenvectors $(i=2,3,4)$ and the non-degenerate state $(j=1)$
\begin{align}
    k \vec{v}_{i}=\vec{e}_{i}^{T} \left(M_{ab}^{N_f=2}\bigg|_{\mathcal{0}(k)}-M^{k0} \right) \vec{e}_{1}\;, \qquad k\vec{v}= 
    \begin{pmatrix}
        \frac{ie q_dq_uC\sqrt{2}\kt\cdot\Bt}{\chi\sqrt{q_d^2+q_u^2}}\\ \frac{ie(q_d^2-q_u^2)C\kt\cdot\Bt}{2\chi\sqrt{q_d^2+q_u^2}} \\
        \frac{ie(q_d^2-q_u^2)C\kt\cdot\Bt}{2\chi\sqrt{q_d^2+q_u^2}}
    \end{pmatrix}.
\end{align}
From the diagonal components of the matrix in \eqref{eq:diagpmat}, we immediately obtain the first-order corrections to the eigenvalues,
\begin{align}
\lambda_2'&= 0\nonumber,\\
\lambda_3'&= -\frac{ieC\kt\cdot\Bt}{\chi\sqrt{2}}\sqrt{q_d^2+q_u^2},\nonumber\\
\lambda_4'&= \frac{ieC\kt\cdot\Bt}{\chi\sqrt{2}}\sqrt{q_d^2+q_u^2}.
\label{eq:EValCorrections}
\end{align}
Our shifted eigenvalues are $\lambda_{i,{\rm tot}}=\lambda_i+\lambda_{i}'$, and from the relation $\omega=i\lambda$ we obtain the shifted frequencies,
\begin{align}
\omega_{2,{\rm new}} &= 0,  \\
\omega_{3,{\rm new}} &=  -\frac{eC}{\chi\sqrt{2}}\sqrt{q_d^2+q_u^2}|\kt\cdot\Bt|, \\
\omega_{4,{\rm new}} &= \frac{eC}{\chi\sqrt{2}}\sqrt{q_d^2+q_u^2}|\kt\cdot\Bt|.  
\end{align}

The first-order corrections to the eigenvectors take the form
\begin{align}
    \et_{1,{\rm new}}&=\et_{1}+\frac{k}{2\gamma}\vec{v}_{i} \et_{i}\;, \\
    \et_{i,{\rm new}}&=\et_{i} -\frac{k}{2\gamma}  \vec{v}_{i} \et_{1}, 
\end{align}
for $i=2,3,4$. Using this prescription, we compute the shifted eigenvectors,
\begin{align}
    \et_{1,{\rm new}}&= \left\{\frac{ieCq_u\kt\cdot\Bt}{\gsph\chi 2\sqrt{2}},\frac{1}{\sqrt{2}},\frac{ieCq_d\kt\cdot\Bt}{\gsph\chi 2\sqrt{2}}, \frac{1}{\sqrt{2}}\right\},\\
    \et_{2,{\rm new}}&= \left\{\frac{q_d}{\sqrt{q_d^2+q_u^2}},-\frac{ieCq_dq_u\kt\cdot\Bt}{\chi\gsph2\sqrt{q_d^2+q_u^2}},\frac{q_u}{\sqrt{q_d^2+q_u^2}}, -\frac{ieCq_dq_u\kt\cdot\Bt}{\chi\gsph2\sqrt{q_d^2+q_u^2}} \right\},\\
    \et_{3,{\rm new}}&= \left\{-\frac{q_u}{\sqrt{2(q_d^2+q_u^2)}}, \frac{1}{8}\left(4-\frac{ieC\sqrt{2}\kt\cdot\Bt(q_d^2-q_u^2)}{\chi\gsph\sqrt{q_d^2+q_u^2}}\right) ,\frac{q_d}{\sqrt{2(q_d^2+q_u^2)}},-\frac{1}{8}\left(4+\frac{ieC\sqrt{2}\kt\cdot\Bt(q_d^2-q_u^2)}{\chi\gsph\sqrt{q_d^2+q_u^2}}\right)\right\},\\
    \et_{4,{\rm new}}&= \left\{-\frac{q_u}{\sqrt{2(q_d^2+q_u^2)}}, -\frac{1}{8}\left(4+\frac{ieC\sqrt{2}\kt\cdot\Bt(q_d^2-q_u^2)}{\chi\gsph\sqrt{q_d^2+q_u^2}}\right) ,\frac{q_d}{\sqrt{2(q_d^2+q_u^2)}},\frac{1}{8}\left(4-\frac{ieC\sqrt{2}\kt\cdot\Bt(q_d^2-q_u^2)}{\chi\gsph\sqrt{q_d^2+q_u^2}}\right)\right\}.
\end{align}

\end{widetext}

\end{document}